\documentclass[%
 aip,
 amsmath,amssymb,
 reprint,%
]{revtex4-1}

\usepackage{graphicx}
\usepackage{dcolumn}
\usepackage{bm}

\usepackage[utf8]{inputenc}
\usepackage[T1]{fontenc}
\DeclareUnicodeCharacter{00A0}{ }
\usepackage{mathptmx}
\usepackage{etoolbox}
\usepackage{chemformula}
\usepackage{physics}
\usepackage{color,soul}
\usepackage{float}
\usepackage{hyperref}

\newcommand{\bmvop}{\bm{\hat{v}}}
\newcommand{\vop}{\hat{v}}
\newcommand{\bmxop}{\bm{\hat{x}}}
\newcommand{\bmaop}{\bm{\hat{A}}}
\newcommand{\bmeop}{\bm{\hat{E}}}
\newcommand{\oprho}{\hat{\rho}}
\newcommand{\bmk}{\bm{k}}
\newcommand{\bmq}{\bm{q}}
\newcommand{\bmr}{\bm{r}}
\newcommand{\bmrr}{\bm{R}}

\newcommand{\calla}{\mathcal{A}}

\renewcommand{\braket}[1]{\expval{#1}}

\newcommand{\anni}{\hat{c}}
\newcommand{\crea}{\hat{c}^{\dagger}}
\newcommand{\annid}{\hat{d}}
\newcommand{\cread}{\hat{d}^{\dagger}}

\makeatletter
\def\@email#1#2{%
 \endgroup
 \patchcmd{\titleblock@produce}
  {\frontmatter@RRAPformat}
  {\frontmatter@RRAPformat{\produce@RRAP{*#1\href{mailto:#2}{#2}}}\frontmatter@RRAPformat}
  {}{}
}%
\makeatother
\begin{document}

\preprint{AIP/123-QED}

\title{Recent Progress in the Theory of Bulk Photovoltaic Effect}
\author{Zhenbang Dai}

\author{Andrew M. Rappe}%
\email{rappe@sas.upenn.edu}
\affiliation{ 
Department of Chemistry, University of Pennsylvania, Philadelphia, Pennsylvania 19104–6323, USA
}%

\date{\today}

\begin{abstract}
(The following article has been submitted to by Chemical Physics Reviews. After it is published, it will be found at \href{https://publishing.aip.org/resources/librarians/products/journals/}{Link}.)
\\
The bulk photovoltaic effect (BPVE) occurs in solids with broken inversion symmetry and refers to  DC current generation due to uniform illumination, without the need of heterostructures or interfaces, a feature that is distinct from the traditional photovoltaic effect.
Its existence has been demonstrated almost 50 years ago, but predictive theories only appeared in the last ten years, allowing for the identification of different mechanisms and the determination of their relative importance in real materials.
It is now generally accepted that there is an intrinsic mechanism that is insensitive to scattering, called shift current, where first-principles calculations can now give highly accurate predictions. 
Another important but more extrinsic mechanism, called ballistic current, is also attracting a lot of attention, but due to the complicated scattering processes, its numerical calculation for real materials is only made possible quite recently. 
In addition, an intrinsic ballistic current, usually referred to as injection current, will appear under circularly-polarized light and has wide application in experiments. 
In this article, experiments that are pertinent to the theory development are reviewed, and a significant portion is devoted to discussing the recent progress in the theories of BPVE and their numerical implementations.
As a demonstration of the capability of the newly developed theories, a brief review of the materials design strategies enabled by the theory development is given.  
Finally, remaining questions in the BPVE field and possible future directions are discussed to inspire further investigations. 

\end{abstract}

\maketitle

\begin{quotation}
\tableofcontents
\end{quotation}

\section{\label{sec:level1}Introduction}
The bulk photovoltaic effect (BPVE), sometimes also called the photogalvanic effect (PGE), refers to the electric current generation in a homogeneous material under light illumination, in contrast to the traditional photovoltaics where a heterojunction, such as a p-n junction, is needed to separate the photo-generated carriers \cite{Fridkin01p654, vonBaltz81p5590, Sipe00p5337, Young12p116601}. 
It has attracted increasing interest among the communities of material science and material engineering due to its potential to surpass the Shockley-Queisser limit governing traditional solar cells \cite{shockley1953electrons, Spanier16p611}, and the simplified device geometry due to the homogeneity is also promising for fabricating light detectors.
Meanwhile, intense theoretical investigations have been made to understand the physical origin of BPVE\cite{Tan16p16026, Morimoto16pe1501524}, which seems to be a rather peculiar phenomenon, as one would naively expect that in the absence of  built-in field (as exists in a p-n junction), the oscillating light field would only drive the charge carriers to oscillate periodically without inducing a net current.
Thus, the breaking of such intuition implies that we must go beyond the linear response regime to fully account for the BPVE. 

\begin{figure}
    \centering
    \includegraphics[width=60mm]{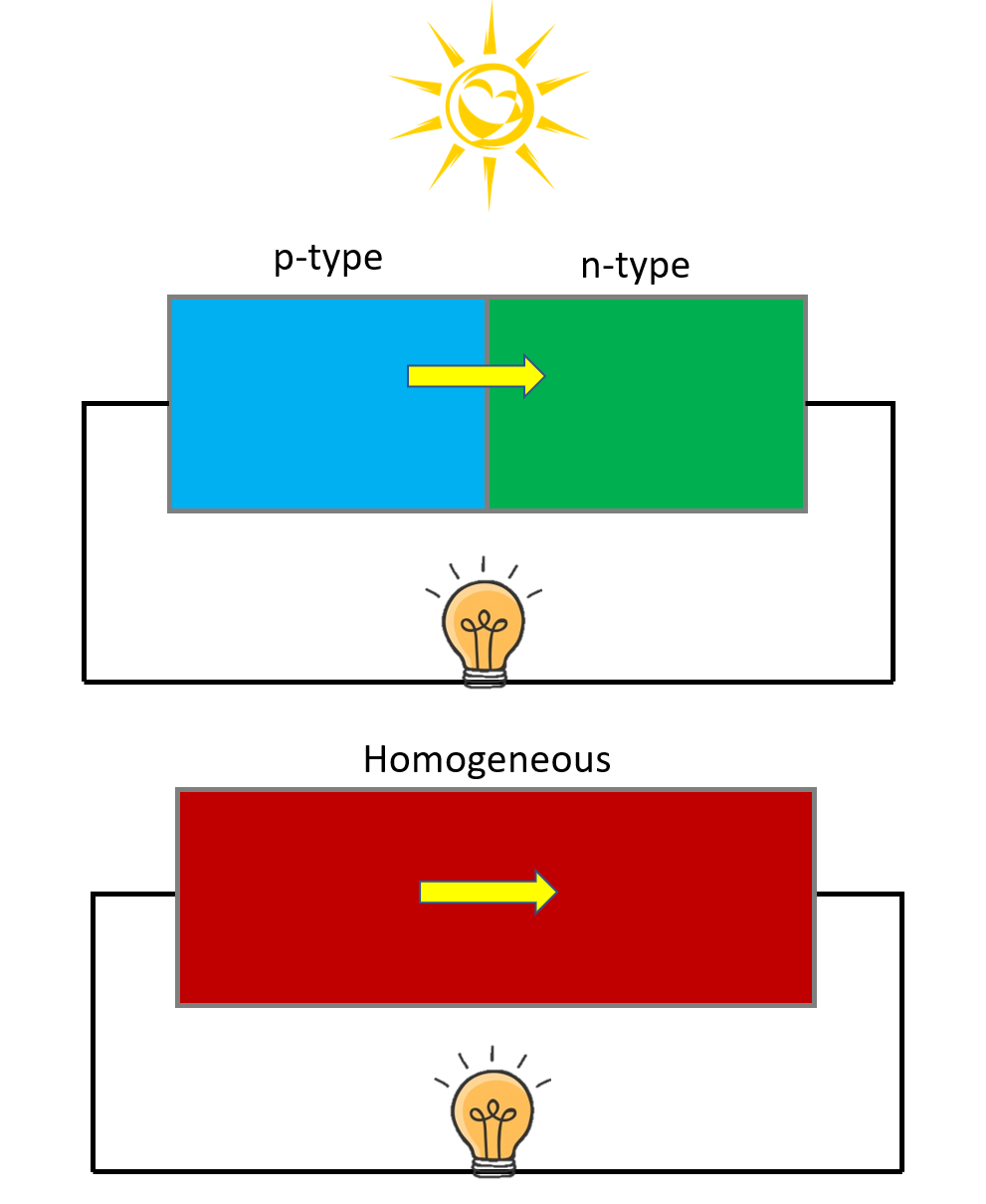}
    \caption{Illustration of traditional photovoltaics and bulk photovoltaics.
    For traditional photovoltaics (upper panel), a heterojunction is needed where the built-in electric field can separate the photoexcited carriers. 
    The bulk photovoltaic effect (lower panel), on the other hand, can occur in a homogeneous material lacking inversion symmetry.}
    \label{fig:pn_vs_bpve}
\end{figure}

To our knowledge, the first investigations of BPVE were conducted in the late 1960s and early 1970s.  Experiments were  carried out to measure the BPVE photocurrent for several materials that do not have an inversion center, and  theories based on time-dependent perturbation theory were formulated for different mechanisms \cite{Belinicher82p649, Fridkin74p433,Fridkin93p1}. 
A review of the research work at that time can be found in the book by Sturman and Fridkin  \cite{Fridkin01p654}.
These early works contribute significantly to the understanding of BPVE, where important concepts such as \textit{shift current} and \textit{ballistic current} were introduced as possible mechanisms for the DC photocurrent, but the simplified models employed by those works hindered the further interpretation and prediction of this effect for real materials in a broader spectral range. 
{
Progress was made toward calculating BPVE in real materials by adopting different electronic structure theories \cite{Hornung83p225, Nastos06p035201, Nastos10p235204, Kral00p1512}, but a truly first-principles calculation with direct comparison with experiments had been lacking.}
In 2012, Young and Rappe revisited the theories of BPVE and adapted the \textit{shift current} theory into a formula that is amenable to first principles calculations \cite{Young12p116601}. 
The great agreement between the first-principles simulations and the experimental results for \ch{BaTiO3} reinvigorated the theoretical study of BPVE due to the demonstrated predictive power of first-principles calculations \cite{Ibanez18p245143, Wang17p115147}.
Later, first-principles models for other BPVE mechanisms have been reported \cite{Dai21p177403, Dai21p235203}, which in general improve the accuracy of the theoretical prediction, and different design routes for enhancing the BPVE in materials are suggested based on the developed first-principles calculations \cite{Cook17p14176, Wang16p10419, Schankler21p1244}.
Besides the exciting progress in numerically calculating BPVE, there is also a renewed theory development that is mainly based on Floquet theory and non-equilibrium Green's function (NEGF) formalism instead of the well-developed perturbation theory \cite{Morimoto16pe1501524, Ishizuka17p033015, Bajpai19p025004}. 
These newly developed theories allow for the study of BPVE for finite systems and their temporal and spatial behavior. 
Also, the topological nature of BPVE is explored by several works which propose that BPVE can serve as a way to probe the topological phase transition \cite{Morimoto16pe1501524, Juan17p15995}. 

In this review article, we  survey the recent progress in theories and numerical calculations in the field of the bulk photovoltaic effect, aiming to introduce the basic concepts as well as the latest understanding of this effect. 
The rest of this article is organized as follow: In Section~\ref{sec:rev_exp}, we will have a brief review of some early and recent experiments measuring BPVE and the characteristics of the observed photocurrent, which are pertinent to the development of the BPVE theory. 
Then, in Section~\ref{sec:theory}, we start to discuss the theories proposed for BPVE, including the theory of shift current, ballistic current, and injection current.
The recently developed Floquet and NEGF formalisms will also be introduced there. 
We will present the numerical implementations for each mechanism and talk about the technical details. 
Following the development of BPVE theory, in Section~\ref{sec:mat_design} we will go over the strategies of improving BPVE in real materials that are guided by the theory framework introduced in the previous sections.
In the last section, we will summarize the content introduced in this article and give our perspective on the future development of the bulk photovoltaic effect.


\section{Review of Experiments}
\label{sec:rev_exp}
Although there is no consensus on which experiment observed the BPVE for the first time, it is clear that in the 1970s numerous experiments demonstrated the existence of a steady DC current in homogeneous materials \textit{lacking inversion symmetry} uniform illumination. These observations inspired a plethora of theoretical studies to understand this phenomenon \cite{Fridkin01p654}.
More recently, there have been experiments trying to clarify the nature of the observed photocurrent, with a focus on trying to separate the contributions from different mechanisms \cite{Burger19p5588, Burger20p081113}. 
Meanwhile, it has been shown experimentally by Spanier {\em et al}.\cite{Spanier16p611} that the BPVE can indeed surpass the Shockley-Queisser limit. 
Therefore, to understand the signature and significance of BPVE, we devote this section to reviewing some of the important experiments.

\subsection{\label{sec:koch_exp} BPVE of tetragonal \ch{BaTiO3}}

\begin{figure}
    \centering
    \includegraphics[width=0.45\textwidth]{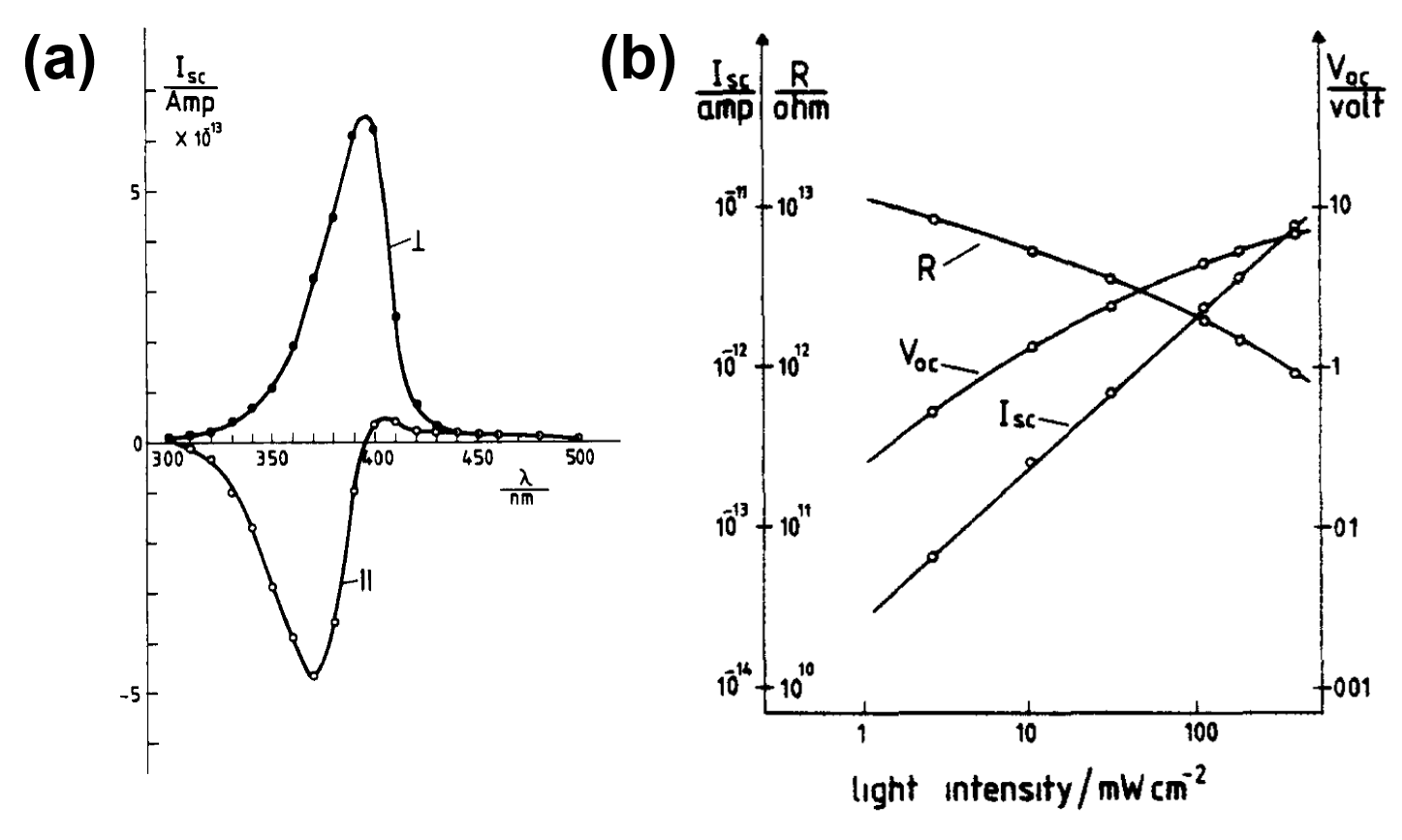}
    \caption{Polarization and temperature dependence of BPVE in \ch{BaTiO3}. (a) $zzZ (\parallel$) and $xxZ (\perp$) components of the photocurrent in \ch{BaTiO3} and their frequency (wavelength) dependence. The lower-case letters represent the direction of light polarization, and the upper-case letters represent the direction of photocurrent. 
    The photocurrent shows a clear polarization dependence and frequency dependence.
    (b)The intensity dependence of open-circuit voltage, short-circuit current, and sample resistance of \ch{BaTiO3} at 488~nm wavelength. 
    Of particular importance is the linear-dependence of the short-circuit current.
    Reproduced from the work\cite{Koch75p847}.}
    \label{fig:kohc_exp}
\end{figure}

{The signatures of BPVE perhaps were detected as early as 1930s, where the photoelectret effect was observed in ferroelectric materials \cite{Nadjakoff37p1865, Nadjakoff38p226}. 
In photoelectrets, the light illumination could induce a long-lasting change in the polarization, and a built-in electric field can be observed.
This polarization difference between the ground state and excited states may share the same origin with the BPVE.
Later, a large open-circuit voltage (${V_{oc}}$) induced by light was found in \ch{SbSI_{0.35}Br_{0.65}} and \ch{BaTiO3} between 1970 and 1972 \cite{Grekov70p423, Volk72p3214}, which strongly indicated the existence of BPVE in these ferroelectric materials, as traditional photovoltaic effects cannot have ${V_{oc}}$ larger than the band gap.}
In 1974, Glass \cite{Glass74p233} provided the concrete evidence of BPVE by showing a steady-state DC current in iron-doped \ch{LiNbO3} and  linear scaling with  light intensity.
Although these early works are undoubtedly crucial for BPVE, the most important experiments in terms of recent theory development can be argued to be the ones conducted by Koch {\em et al}.\ in 1975 and 1976 for \ch{BaTiO3}, where the  dependence of the photocurrent on light polarization, frequency, and intensity were clearly reported \cite{Koch75p847, Koch76p305}. It is this set of experiments that most of recent first-principles simulations compare with \cite{Young12p116601,Dai21p177403}.  


To measure the bulk photovoltaic properties of \ch{BaTiO3}, several single-crystal samples were fabricated and poled with an electric field about 5~kV/cm while cooling down through the Curie temperature to ensure uniformity. 
The samples were usually  0.02-0.05~cm thick, 0.1--0.2~cm wide, and 0.1--0.3~cm long.
Then, a 488~nm laser focused to 0.03~cm diameter was scanned through the sample from one electrode to the other. 
What is significant is that the open-circuit voltage was non-zero, larger than the band gap of \ch{BaTiO3}, and almost uniform across the sample, except when quite near to the electrode. This is different from the traditional photovoltaic effect, where the charge carrier separation can only happen at an interface of distinct materials, and the open-circuit voltage is usually smaller than the band gap. 
This position-independent behavior was largely unexplored until a recent study \cite{Ishizuka17p033015} (See Section.~\ref{sec:theory_finte_sys}) in which the real-space distribution of BPVE was simulated explicitly. 

More importantly, the spectral dependence of the open-circuit voltage (or equivalently the short-circuit current \cite{Koch76p305}) was measured in a range of photon energies above the band gap of \ch{BaTiO3}, as shown in Fig.~\ref{fig:kohc_exp}(a).
 Two orthogonal linear polarizations of light were used, and distinct spectral behaviors were obtained. 
In particular, the $zzZ$ component (lower-case letters representing the light polarization and  upper-case letter representing the current direction) exhibited a sign change around 390~nm. 
When changing the intensity of light, a linear scaling was found, showing that the observed photocurrent is a second-order response to electric field.
This combination of behaviors is unique to BPVE and cannot be understood by traditional photovoltaic effect, so any theories trying to explain BPVE are built on these observations.

\subsection{\label{sec:separate_sc_bc_exp}Separation of Different Mechanisms}

\begin{figure}
    \centering
    \includegraphics[width=0.45\textwidth]{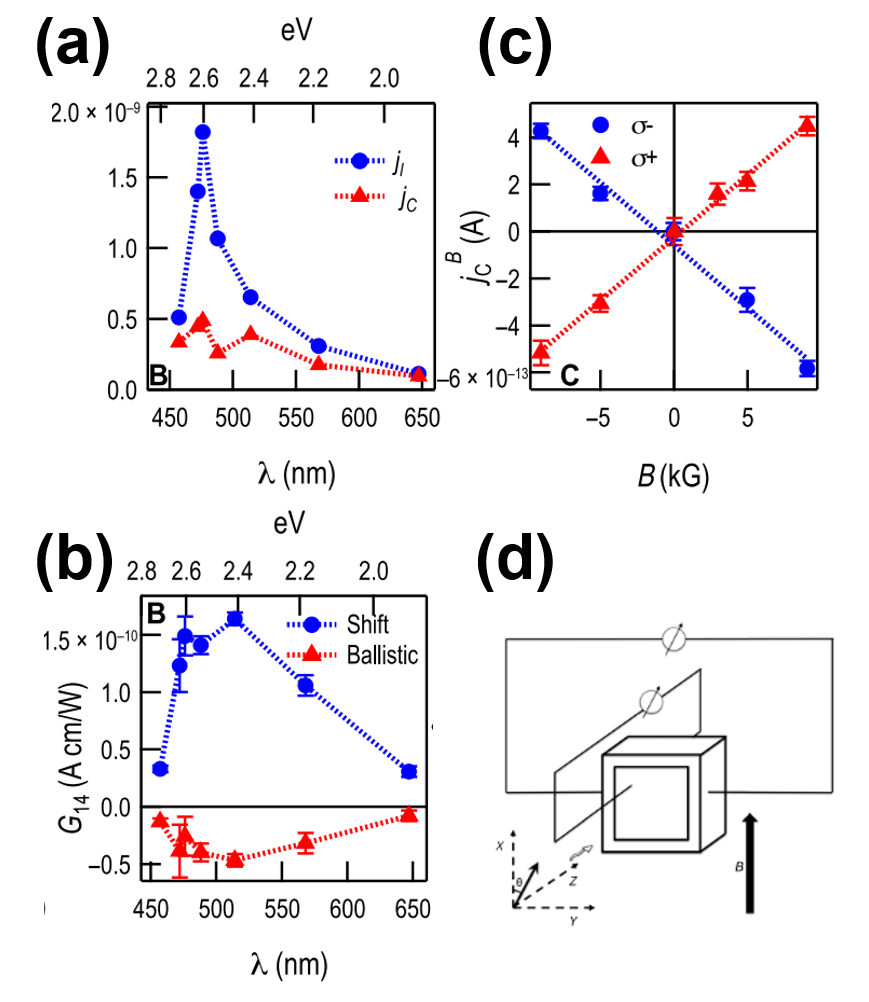}
    \caption{Separation of shift current and ballistic current. All the data showing here are for \ch{Bi12GeO20}.
    (a) The wavelength dependence of photocurrents under linearly- and circularly-polarized light.
    (b) Hall current for circular photocurrents under different illumination helicity.
    (c) Separation of ballistic current and shift current. This separation is based on the assumption that only ballistic current will have Hall effect and that it is Lorentz-like.
    (d) The illustration of experimental setup for separating the ballistic and shift current.
    Reproduced from the work \cite{Burger19p5588}.}
    \label{fig:sep_sc_bc_exp}
\end{figure}

Since the early experiments were conducted showing the existence of BPVE in non-centrosymmetric (breaking P-symmetry)
materials, including the ones by Koch\cite{Koch75p847, Koch76p305}, two major complementary theories were proposed to explain the observed photocurrent, namely the \textit{shift current} and \textit{ballistic current}.
A detailed introduction and discussion of these can be found in Sec.~\ref{sec:theory}, but for the purpose of illustrating the ideas underlying the experiments discussed here, it suffices to know that it is believed that the shift current will be less susceptible to magnetic field whereas the ballistic current can give rise to a Hall current as any classical charge current does \cite{Ivchenko84p55}. 
Therefore, to validate the shift current and ballistic current theory, people have designed experiments trying to separate the two types of current with the help of a uniform electric field.

An outstanding work was conducted by Burger {\em et al}. \cite{Burger19p5588} in which \ch{Bi12GeO12} was chosen as the target material. 
It belongs to the $T$ space group (\#197), which dictates that only the $yzX$  component of the photocurrent (and all permutations of the indices) could have non-vanishing value for linearly-polarized light. 
For circularly-polarized light, a similar symmetry argument shows that only $xX$, $yY$, and $zZ$ will be non-zero, with lower-case letter showing the propagation direction of the light. 
These symmetry properties are particularly  useful, since in the absence of magnetic field, a linear light whose polarization lies within $yz$-plane or circular light that is propagating along the $x$-axis could only generate photocurrent flowing along the $x$-direction, so that a Hall current after turning on a magnetic field along $y$-axis can be uniquely identified along the $z$-direction. 
If, however, the system already had non-magnetic BPVE current along directions other than $x$-axis, then extra effort and caution would have to be taken to separate the Hall current from the ``intrinsic'' response.

Their procedure for the current separation is as follows: under  circularly-polarized light, there is only photocurrent of ballistic type (See Section~\ref{sec:theory}), so one can extract the mobility of the carriers $\mu_{nth}$ of ballistic current via Hall effect.
For linearly-polarized light, however, there could exist both shift current and ballistic currents, and only the ballistic current is believed to  respond to magnetic field. 
So, one can apply a magnetic field again, obtaining the Hall current under linear light, and then calculate the non-magnetic ballistic current with the help of $\mu_{nth}$. 
After subtracting the non-magnetic ballistic current from the total non-magnetic current, the non-magnetic shift current can be acquired, and  the contributions from these two mechanisms can thus be separated.

Though such procedure looks reasonable, there are two caveats. 
For one, this experiment is designed under the assumption that shift current does not respond to static and uniform magnetic field.
This was firstly argued by Ivchenko \cite{Ivchenko84p55} in which he stated that as long as the cyclotron frequency of magnetic field is much smaller than the difference between the light frequency and band gap, then the shift current will be barely impacted by the magnetic field, without further proof. 
Thus, the validity of this assumption remains to be examined. 
For another, it is worth mentioning that in the same paper by Ivchenko, the authors demonstrated that the magnetic field can break the time-reversal symmetry and induce a new current, which will be proportional to the non-magnetic shift current for a two-band model.
So, in another work by Burger \cite{Burger20p081113} {\em et al}., they took the new current into account and discussed several different scenarios for the relative magnitude of this current, leaving the exact separation of ballistic current and shift current still an open question.
Despite these caveats, this work constitutes an important step forward toward experimentally verifying various BPVE mechanisms. 

\subsection{\label{sec:exceed_sq_limit}Beyond Shockley-Queisser limit}

\begin{figure}
    \centering
    \includegraphics[width=0.45\textwidth]{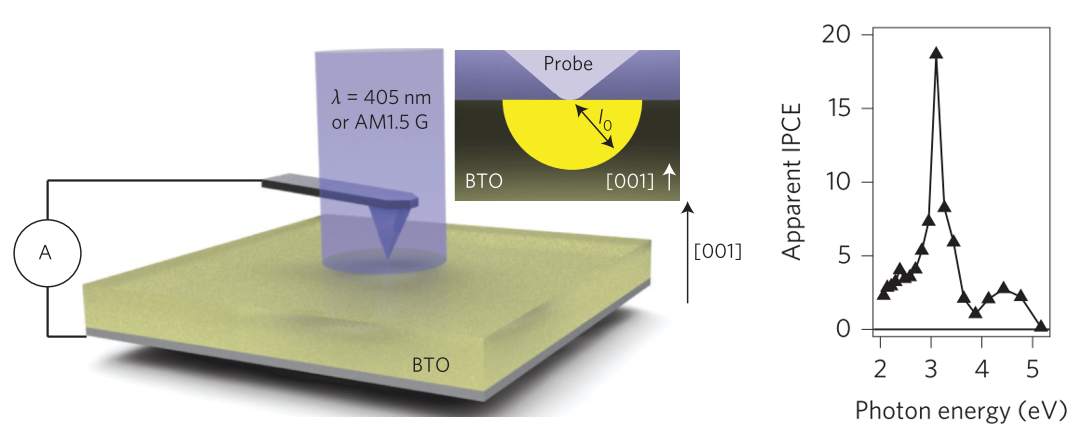}
    \caption{Experimental schematic for demonstrating  power conversion efficiency  above the Shockley-Queisser limit.
    The electric field within the region around the tip can enhance the carrier multiplication so that the incident photon-to-collected electron (IPCE) efficiency can exceed unity.
    Reproduced with permission from Nat. Photonics 10, 616 (2016). \cite{Spanier16p611} Copyright 2016 Macmillan Publishers Limited, part of Springer Nature. All rights reserved.}
    \label{fig:more_sq}
\end{figure}

In addition to the experiments designed to understand the fundamental physics of the BPVE, there is also great research interest toward exploiting the BPVE in real-world applications \cite{Peng20p3933, Perez19p100350, Wang20p126831}. 
Since the BPVE is not governed by the rules of traditional photovoltaics, it is in principle not limited by the Shockley-Queisser limit that is imposed on traditional solar cells. 

The Shockley-Queisser limit explains and quantifies that for high efficiency, the ideal band gap for a traditional solar should not be too large or too small \cite{shockley1953electrons}.
If the band gap is too large, then a large portion of the sunlight spectrum is unable to be absorbed. If the band gap is too small, then even though more sunlight can be absorbed, the photoexcited electrons will initially occupy higher-energy states but will rapidly thermalize and relax to the conduction band bottom before they can be harvested by the electrodes. 
Thus, for the solar spectrum, there exists a perfect band gap that  maximizes the power conversion efficiency, and for any specific band gap value, there exists a maximum power conversion efficiency. 
However, for the bulk photovoltaic effect, both shift current and ballistic current mechanisms involve non-thermalized carriers  giving rise to a current (See Sec.~\ref{sec:theory}), so the Shockley-Queisser limit no longer applies. 

To demonstrate the capability of the BPVE to achieve  high power conversion efficiency, Spanier {\em et al}.\ \cite{Spanier16p611} employed a tip-enhanced geometry\cite{Alexe11p246}
that can effectively harvest  the non-thermalized electrons in \ch{BaTiO3}. 
Such geometry makes use of the fact that the electrode tip can screen the polarization bound charge in a very confined region so that it can create a very large electric field around the tip, which will let the non-thermalized electrons ionize more electrons from valence bands and effectively generate an incident photon-to-collected electron (IPCE) efficiency larger than unity. 
This process should be distinguished from charge separation caused by the band bending at the ferroelectric-electrode interface that falls into the category of traditional photovoltaics. 
Rather, it manifests an efficient usage of the hot carriers with already asymmetrically distributed momenta that are caused by the BPVE (more specifically, the ballistic current mechanism). 
As a result, the power conversion efficiency of \ch{BaTiO3} for this geometry is 4.8\%, around 50\% higher than the Shockley-Queisser limit for materials with a 3.2~eV band gap, which shows the potential of using BPVE to design next-generation high-efficiency photovoltaic devices.

\section{Theory development and Numerical Implementation}
\label{sec:theory}
Over the past 50 years, different theories have been proposed to understand the nature of BPVE, and most of them are based on the time-dependent perturbation theory, either in density matrix form \cite{Kraut79p1548, vonBaltz81p5590, Aversa95p14636} or in second-quantized form \cite{Sipe00p5337, Parker19p045121}. 
These theories are especially successful in explaining the DC photocurrent for bulk materials under linearly-polarized or circularly-polarized light, but they do not address the temporal response or account for spatially inhomogeneous  light illumination. 
Thus, in recent years, Floquet theory combined with non-equilibrium Greens function (NEGF) methods have been developed aiming to address these issues \cite{Morimoto16pe1501524, Ishizuka17p033015, Bajpai19p025004, Ishizuka21p10}.
In this section, we will first review the original theories of BPVE based on perturbation theory, and then the Floquet theory and NEGF methods will be discussed at some length to provide perspective on their advantages and shortcomings. 

\subsection{\label{sec:theory_linear_light} Linearly-Polarized Light}
Experimentally, the scaling of BPVE photocurrent with the light intensity is linear \cite{Koch75p847}, so for linearly-polarized light, a phenomenological description \cite{Fridkin01p654, Hornung21p195203} of BPVE can be written as
\begin{align}
\label{eqn:phenom}
    j^q = {\sigma}_{rs}^{q} E_{r} E_{s},
\end{align}
where $j^q$ is the photocurrent density along the Cartesian direction $q$, $E_r$ and $E_s$ are the components of the electric field of light, and ${\sigma}_{rs}^{q}$ is the response tensor or susceptibility tensor that characterizes the BPVE for a certain system.
There are some other conventions of labeling the response tensor in the literature, such as ${\sigma}_{rsq}$ or ${\sigma}_{qrs}$, where $q$ represents the current propagation direction. 
In this review, we will stick with the conventions that $q$ always appears in the superscript or in the last place.
As $E_rE_s$ is proportional to the light intensity, this expression can correctly describe the scaling behavior of BPVE with light intensity.
It is noted that such second-order response has already imposed the symmetry constraint that only non-centrosymmetric (no inversion center) structures can have BPVE. 
To see this, imagine applying an inversion operation to the system. Polar vectors such as $j^q$, $E_r$, and $E_s$ will acquire a minus sign:
\begin{align*}
    j^q &\rightarrow -j^q, \\
    E_r &\rightarrow -E_r, \\
    E_s &\rightarrow -E_s.
\end{align*}
If the system possesses inversion symmetry, then the response tensor ${\sigma}_{rs}^{q}$, an intrinsic property of the material, will return to itself after the inversion operation. 
As a result, inversion symmetry results:
\begin{align*}
    -j^q &= {\sigma}_{rs}^{q}(-E_r)(-E_s)
    ={\sigma}_{rs}^{q} E_r E_s \\
    &=j^q,
\end{align*}
which indicates that $j^q$ will always be zero for a centrosymmetric structure.
Therefore, breaking inversion symmetry is a prerequisite for BPVE, and accordingly BPVE can be used in detecting phase transitions involving  inversion-symmetry breaking \cite{Ji19p955}.

To develop a microscopic theory for ${\sigma}_{rs}^{q}$, one can start with a non-interacting many-body system where the two-body interaction is effectively treated in a mean-field fashion, a strategy that is widely used in modern electronic structure calculations such as density functional theory (DFT) and the Hartree-Fock approximation \cite{Giuliani05quantum}.
Then, one is interested in how the the equilibrium density matrix of this system will evolve under the perturbation from light. 
We would especially like to know the form of the resulting non-equilibrium steady-state density matrix.
More concretely, under the dipole approximation, the electron-light interaction in the velocity gauge \cite{Bandrauk13p153001, Jishi13Feynman} can be expressed as the following minimal coupling form:


\begin{align}
    \label{eqn:e-light}
    \hat{H}_{e-{\rm light}} = e\bmvop \cdot \bmaop(t) 
    = e\bmvop \cdot \bmaop_0 (e^{i\omega t}+e^{-i\omega t}), 
\end{align}
and the full Hamiltonian can be written as:
\begin{align}
    \label{eqn:full_H}
    \hat{H} = \hat{H}_0 + \hat{H}_{e-{\rm light}}, \\
    \hat{H}_0 = \sum_{n} \varepsilon_{n} 
    \ket{\psi_{n}} \bra{\psi_{n}}.
\end{align}
Here, $\bmvop$ is the velocity operator and $\bmaop$ is the vector potential of light, which can be rewritten as $\bmaop(t) = -\frac{i}{\omega}\bmeop(t)$ in the velocity gauge \cite{Jishi13Feynman}. 
$\hat{H}_0$ describes the non-interacting Hamiltonian with known energy spectrum $\varepsilon_{n}$ and eigenstates $\ket{\psi_{n}}$.
Then, the equilibrium (non-perturbed) density matrix (operator) can be constructed as:


\begin{align}
    \label{eqn:rho0}
    \oprho_0 = \sum_{i} 
    p_{i}
    \ket{\psi_{i}} \bra{\psi_{i}},
\end{align}
where $p_i$ is the probability of being in the many-body state $i$.
We would like to know the steady-state density matrix $\oprho(t)$ under continuous illumination because the steady-state current can be computed as:
\begin{align}
    \label{eqn:current_density_op}
    j^q = Tr[\oprho^I(t) \bmvop^I (t)] 
    = \sum_{nm} \rho^I_{nm}(t) v^I_{mn} (t).
\end{align}
Note that the form of Eq.~(\ref{eqn:current_density_op}) is written in the interaction picture, and in this picture, $\oprho^I(t)$ can be calculated perturbatively as:
\begin{align}
    \label{eqn:pertur_rho}
    \oprho^I(t) &= \oprho_0 + \frac{i}{\hbar}
    \int_0^{t}[\hat{H}_{e-{\rm light}}^I(t'), \oprho_0]dt' 
    \nonumber \\
    &-\frac{1}{{\hbar}^2}
    \int_0^{t}\int_0^{t'}
    \big[\hat{H}_{e-{\rm light}}^I(t'), [\hat{H}_{e-{\rm light}}^I(t''), \oprho_0]\big]dt'dt'' + ...,
\end{align}
where $\hat{H}_{e-{\rm light}}^I(t)$ is the perturbation in interaction picture \cite{Jishi13Feynman}. 
As in BPVE theory the response is second-order, we only retain the third term in Eq.~(\ref{eqn:pertur_rho}), and then compute the current $j^q$ via Eq.~(\ref{eqn:current_density_op}). 
After a certain amount of algebra, the steady-state current can be explicitly written  as:
\begin{align}
    \label{eqn:brute_force_j}
    j^q = &\frac{\pi e^3}{ {\omega}^2 }
    \Re \Bigg[\sum_{l,m,n} \sum_{\Omega=\pm \omega}
    \int_{BZ} \frac{d\bmk}{(2\pi)^3} (f_{l\bmk} - f_{n\bmk}) \times
    \nonumber \\
    &\frac{v_{nl}^r(\bmk) v_{lm}^s(\bmk) v_{mn}^q(\bmk)}
    {(\varepsilon_{n\bmk} - \varepsilon_{m\bmk} - i\eta)(\varepsilon_{n\bmk} - \varepsilon_{l\bmk} - \hbar \Omega - i\eta)}
    \Bigg]E_r E_s,
\end{align}
where $f_{l\bmk}$ is the Fermi-Dirac distrubution function, $v_{nl}^r(\bmk) \equiv \braket{n\bmk | \hat{v}^r | l\bmk}$ is the velocity matrix, and $\eta$ is an infinitesimally small value ($0^+$) appearing in the adiabatic turning-on $e^{\eta t}$. \cite{Giuliani05quantum}
Note that we are considering a perfect crystal in the thermodynamic limit, so the dependence of the eigenstates on the crystal momentum $\bmk$ has been made explicit here.

Eq.~(\ref{eqn:brute_force_j}) is one central result for the BPVE theory as it expresses the steady-state current response tensor with quantities that can be obtained from numerical models such as quadratic band structure models, tight binding models or, from first-principles calculations. 
It is therefore tempting to conduct numerical calculations of BPVE based on this expression. However, such calculations would be cumbersome, due to the summation over band index $m$. 
A closer inspection of Eq.~(\ref{eqn:brute_force_j}) will reveal that there is no selection rule for $m$, meaning that in principle one should include an infinite number of bands when summing over $m$.
In practice, even though number of bands in the summation would always be truncated, the long tail due to the function of form $1/ (\varepsilon_{n\bmk} - \varepsilon_{m\bmk} - i\eta)$ will still require a very large number of bands for converged results, which would cause formidable computational cost.
Thus, most numerical calculations will not directly use Eq.~(\ref{eqn:brute_force_j}), but instead employ some further simplified forms.

To simplify the Eq.~(\ref{eqn:brute_force_j}), we will split it into two contributions: the ``three-band'' contribution where $n \neq m$ corresponding to the off-diagonal part of $\oprho(t)$, and ``two-band'' contribution where $n = m$, corresponding to the diagonal part of $\oprho(t)$. 
It turns out that these two contributions will appear under different conditions and thus carry distinct physical meanings.

\subsubsection{Linear Shift Current}
\label{sec:sc}
We will first focus on the three-band contribution, which has a more well-known name, \textit{shift current}. 
The reason why it is called ``shift'' current will become clear later.
After imposing the condition that $n \neq m$, the summation over $m$ can be carried out analytically so as to avoid the necessity of including a large number of bands in numerical calculations. 
The general procedure is to make use of the identity
\begin{align}
    \label{eqn:p_r_commute}
    \bmvop = \frac{i}{\hbar}[\hat{H}_0, \bmxop],
\end{align}
and then replace $v_{mn}^q(\bmk)$ in Eqn.~(\ref{eqn:brute_force_j}). 
The rationale for why it can enable the analytical summation of $m$ is that the Hamiltonian $\hat{H}_0$ in the commutator will give a term $(\varepsilon_{n\bmk} - \varepsilon_{m\bmk})$, which can exactly cancel the principal part in $1/ (\varepsilon_{n\bmk} - \varepsilon_{m\bmk} - i\eta)$.
Care must be taken, however, when summing over $m$ after making this substitution, because now this summation indeed includes the terms $n=m$. 
Thus, we need to manually exclude the terms involving $\braket{n\bmk | [\hat{H}_0, \bmxop] | n\bmk }$.
Now, with the help of the expression for position operator $\bmxop$ in a periodic system by Blount \cite{Blount62p305}, 
\begin{align}
    \label{eqn:pos_blount}
    &\braket{n\bmk' | \bmxop | m\bmk }
    \nonumber \\
    &= -i\delta_{nm} \nabla_{\bmk}\delta(\bmk-\bmk')
    + \delta(\bmk'-\bmk)\braket{u_{n\bmk'} | i\nabla_{\bmk} u_{m\bmk}},
\end{align}
where $u_{n\bmk}$ is the lattice periodic part of the eigenfunction of $\hat{H}_0$, $\psi_{n\bmk}(\bm{x})=e^{i\bmk \cdot \bm{x}} u_{n\bmk} (\bm{x})$,
we can finally rewrite the three-band contribution as:

\begin{align}
    \label{eqn:shift_current}
    j^{q,sh} &= \sigma_{rs}^{q, sh} E_r E_s 
    \nonumber \\
    \sigma_{r s}^{q, sh}&= \frac{\pi e^3}{\omega^2} \sum_{n, l} \int d \mathbf{k}\left(f_{l\bmk}-f_{n\bmk}\right) 
    \braket{n\bmk | \hat{v}^r | l\bmk}
    \braket{l\bmk | \hat{v}^s | n\bmk}
    \nonumber \\
    & \times\left(-\frac{\partial \phi_{n l}^{r}(\mathbf{k}, \mathbf{k})}{\partial k_{q}}-\left[\chi_{l q}(\mathbf{k})-\chi_{n q}(\mathbf{k})\right]\right) 
    \nonumber \\
    & \times \delta\left(\varepsilon_{l}(\mathbf{k})-{\varepsilon_n}(\mathbf{k}) \pm \hbar \omega\right).
\end{align}
Here, $\chi_{n q}(\bmk) \equiv \braket{u_{n\bmk} | i\nabla_{\bmk} u_{n\bmk}}$ is the Berry connection, and $\phi_{n l}^{r}(\mathbf{k}, \mathbf{k})$ is the phase of $\braket{n\bmk | \hat{v}^r | l\bmk}$.
This expression is composed of two parts:
\begin{align}
    \label{eqn:decom_sh}
    \sigma_{r s}^{q, sh}&= e \sum_{n, l} \int d \mathbf{k}
    I_{rs}(n, l; \bmk; \omega) R_{r}^{q}(n, l; \bmk),
\end{align}
with
\begin{align}
    \label{eqn:transition_rate}
    I_{rs}(n, l; \bmk; \omega) 
    &= \frac{\pi e^2}{\omega^2}  \left(f_{l\bmk}-f_{n\bmk}\right) 
    \braket{n\bmk | \hat{v}^r | l\bmk}
    \braket{l\bmk | \hat{v}^s | n\bmk}
    \nonumber \\
    & \times \delta\left(\varepsilon_{l}(\mathbf{k})-{\varepsilon_n}(\mathbf{k}) \pm \hbar \omega\right),
\end{align}
which can regarded as the $\bmk$-resolved transition rate, and 
\begin{align}
    \label{eqn:sv}
    R_{r}^{q}(n, l; \bmk)=
    -\frac{\partial \phi_{n l}^{r}(\mathbf{k}, \mathbf{k})}{\partial k_{q}}-\left[\chi_{l q}(\mathbf{k})-\chi_{n q}(\mathbf{k})\right],
\end{align}
which has a unit of length and can be regarded as the \textit{coordinate shift} of carriers in real-space during the transition.
It is for this reason that Eq.~(\ref{eqn:sv}) is named as shift vector, and the three-band contribution Eq.~(\ref{eqn:shift_current}) is usually called shift current. 
Note that Eq.~(\ref{eqn:shift_current}) is now in a two-band form having only $n$ and $l$, but in essence it is still a three-band expression as the summation of the $m$ is encapsulated in the $k$-derivative terms.
In other words, the first-order expansion of the $k$-derivative terms involves another summation over all states \cite{vonBaltz81p5590}.

Shift current has many interesting properties that are distinguished from classical charge currents. 
For one, it is independent of carrier lifetime and is robust against the scattering by disorder. \cite{vonBaltz81p5590, Morimoto16pe1501524} 
It is not a current carried by classical moving particles as it is exclusively from the coherence of the density matrix, which has no interpretation in the classical picture. 
Instead, it is a manifestation of wave-packet evolution when transitions between different electronic states are happening.
For another, it contains quantum information (the so-called geometrical information) of the electronic structure, as the phases of the wave functions are considered explicitly in the shift vector $R_{r}^{q}(n, l; \bmk)$, whereas for classical charge carriers only group velocities (diagonal elements of the velocity matrix) and occupations (diagonal elements of the density matrix) are relevant. 
Thus, its quantum nature has attracted a lot of attention, and its connection to the modern theory of polarization and topological materials have been explored due to their common relation to Berry connection \cite{Fregoso16preprint}. 

It is now feasible to compute shift current for real materials reliably via first-principles calculations, providing a possible route to quantify its contribution to the experimentally observed photocurrent \cite{Young12p116601, Young12p236601, Ibanez18p245143, Wang17p115147, Brehm18p1470}.
Natos and Sipe demonstrated that the shift current can be calculated from first-principles theories, \cite{Nastos06p035201}, and Young and Rappe revolutionized this field by showing that the first-principles prediction of shift current from density functional theory (DFT) can be directly compared to  experiments \cite{Young12p116601}.
Their formalism bears the caveat that the numerical differentiation of wave functions with respect to $\bm{k}$ might break the gauge invariance (global phase of wave functions) of the shift vector Eq.~(\ref{eqn:sv}). 
Inspired by the strategy employed in the modern theory of polarization \cite{King-Smith93p1651, Vanderbilt00Berry}, the gauge invariance is preserved by transforming the direct derivative into a logarithmic derivative, and the shift currents of tetragonal \ch{BaTiO3} and \ch{PbTiO3} were thus computed using DFT. 

\begin{figure}
    \centering
    \includegraphics[width=0.40\textwidth]{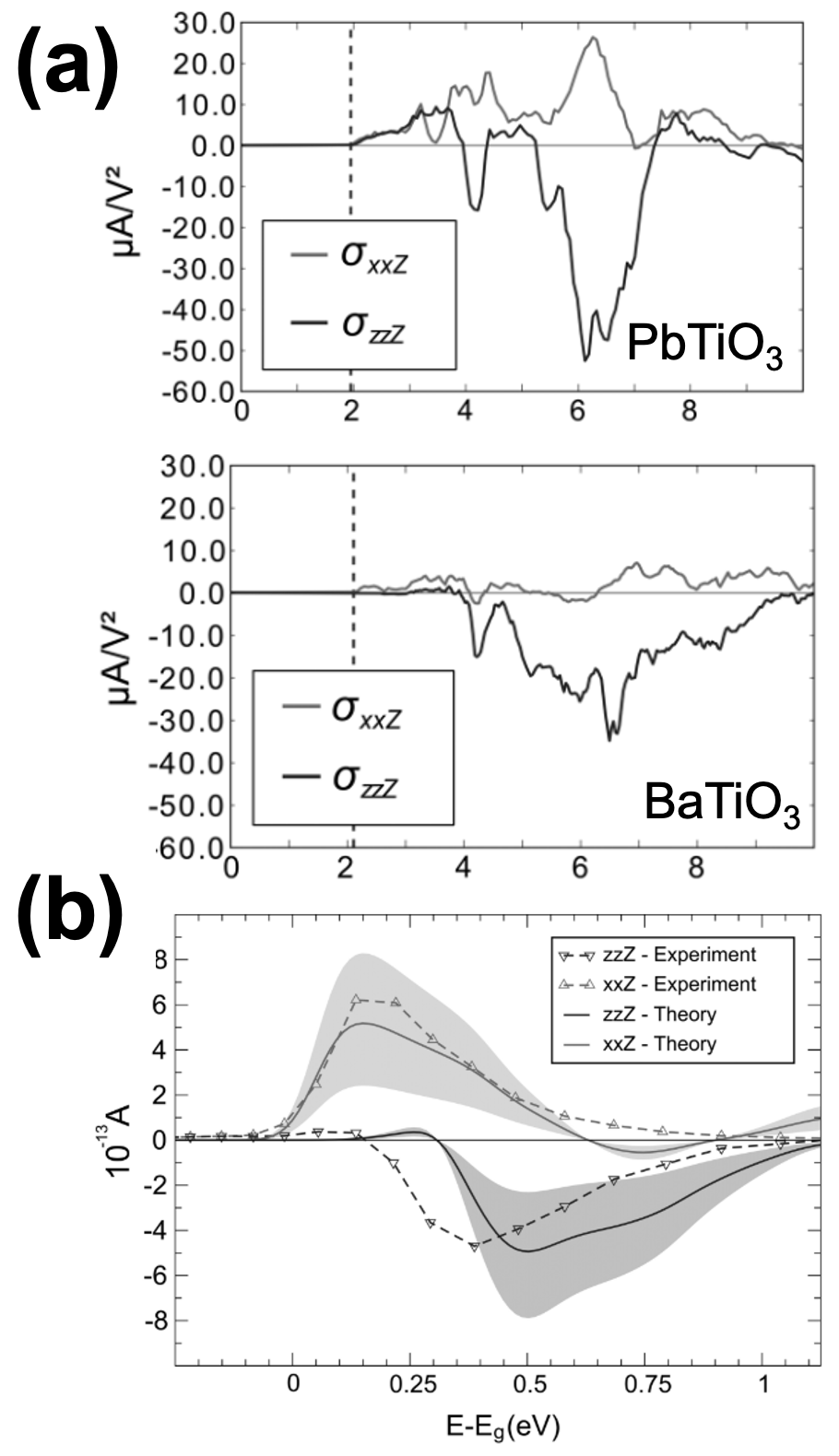}
    \caption{First principles calculations of shift current for \ch{PbTiO3} amd \ch{BaTiO3}. 
    (a) The shift current response tensor for \ch{PbTiO3} (up) and \ch{BaTiO3} (down).
    (b) The actual current computed from first principles and the comparison with experiments. 
    Within the experimental error bar, the computed shift currents at the DFT level has an overall good agreement with experiments.
    Reproduced with permission from Phys. Rev. Lett. 109, 116601 (2012).\cite{Young12p116601} Copyright 2012 American Physical Society.}
    \label{fig:pto_bto_fp}
\end{figure}

Fig.~\ref{fig:pto_bto_fp}(a) shows the $\sigma_{xx}^z$ (gray) and $\sigma_{zz}^z$ (black) components of the current response tensor. 
It is clear that for both \ch{PbTiO3} and \ch{BaTiO3} the largest response is at frequencies much larger than the band edge, which cannot be captured by simple model calculations which usually only consider energy regions around the band edges\cite{Fridkin01p654, Sturman20p407, Shelest79p1353}. 
Also, the shift current calculated from first principles demonstrates large polarization dependence, where $\sigma_{xx}^z$ and $\sigma_{zz}^z$ differ not only in magnitude, but also in sign, which is consistent with experimental observations. 
To make quantitative comparison with experiments by Koch \textit{et al.}\cite{Koch75p847}, Young and Rappe \cite{Young12p116601} also calculated the total shift current flowing through the system by taking into account the absorption of light and the sample dimensions: 
\begin{align}
    \label{eqn:glass_coeff}
    J^{q,sh} = \frac{ {\sigma}_{rs}^{q,sh} }{\alpha_{rs}} E_{r} E_{s} w,
\end{align}
where $J^q$ is total current, ${\alpha_{rs}}$ is the absorption coefficient characterizing how much light can be absorbed and how deep the electric field of light can penetrate, and $w$ is sample width.
In the experiments \cite{Koch75p847, Koch76p305}, the irradiation intensity is $\mathrm{0.35-0.6~mW/cm^2}$, from which the electric field can be deduced, and the sample width is $\mathrm{0.1-0.2~cm}$.
Combined with the theoretical absorption coefficient ${\alpha_{rs}}$, a quantity that is readily evaluated from first principles in the form of Fermi's golden rule \cite{Bassani76p58}, the total shift current can be computed and compared against the experimental photocurrent, as shown in Fig.~\ref{fig:pto_bto_fp}(b). 
What is remarkable is that despite a small mismatch of the frequencies, the calculated shift current of \ch{BaTiO3} can reproduce all the salient features at the band edge, including the overall magnitude, lineshape as well as sign reversal. 
Thus, Young and Rappe inferred that the main contribution to BPVE is shift current, at least for \ch{BaTiO3}.
In addition, shift current is predicted as a mechanism to generate pure spin current (PSC), the first proposal to apply BPVE in spintronic devices.
This will be further discussed in Section.~\ref{sec:linear_inj}.

\begin{figure}
    \centering
    \includegraphics[width=60mm]{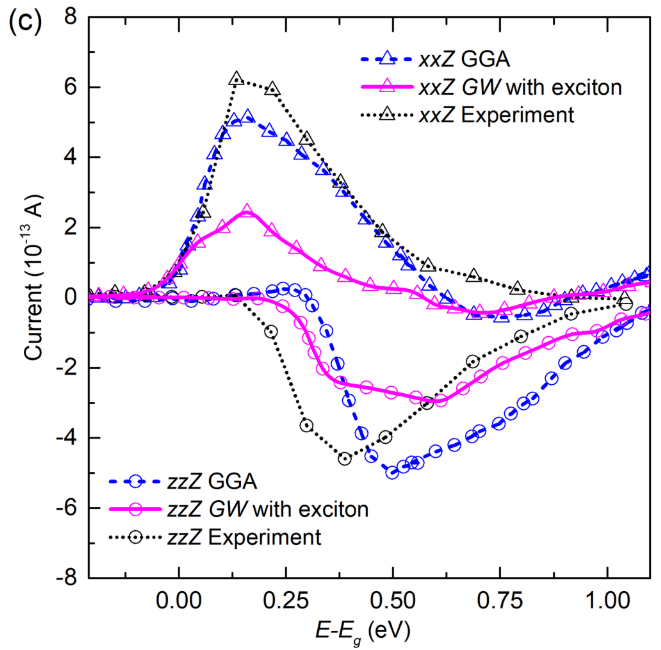}
    \caption{Shift current with exciton correction. By replacing the absorption coefficient computed from DFT with that from Bethe-Salpeter Equation (BSE) and considering the reflectivity, the theoretical shift currents are lowered by half, causing a less satisfactory agreement with experiments.
    Reproduced with permission from Phys. Rev. B 101, 045104 (2020).\cite{Fei20p045104} Copyright 2020 American Physical Society.}
    \label{fig:sc_bse}
\end{figure}

However, a follow-up study by Fei \textit{et al.} \cite{Fei20p045104} shows that the contribution of shift current to BPVE might be exaggerated since the absorption coefficient computed from single-particle approximation (assumed by DFT) will be underestimated. 
After improving the band structure with the $GW$ approximation and introducing the exciton correction to the absorption coefficient \cite{Onida01p601_1}, they found that the calculated shift current will be scaled down such that there is a larger discrepancy between the experimental and theoretical spectra (Fig.~\ref{fig:sc_bse}).
This shows that in addition to the shift current, other mechanisms could also participate in the generation of photocurrent. 
Indeed, as stated earlier, the shift current only originates from the off-diagonal elements of the density operator $\oprho(t)$, but the contribution from the diagonal part, the ``two-band'' contribution, has not been considered. 
Therefore, it remains to examine the contribution from the diagonal part and whether it will improve the BPVE theory. 

\subsubsection{Ballistic Current}
\label{sec:bc}
Under linearly-polarized light, the two-band contribution can be obtained from Eq.~(\ref{eqn:brute_force_j}) by imposing the condition that $n=m$:

\begin{align}
    \label{eqn:inj}
    j^{q,\rm{diag}} &= \frac{\pi \tau_0 e^3}{ {\omega}^2 \hbar }
    \Re \Bigg[\sum_{l,n} \sum_{\Omega=\pm \omega}
    \int_{BZ} \frac{d\bmk}{(2\pi)^3} (f_{l\bmk} - f_{n\bmk}) 
    \nonumber \\
    &\times v_{nl}^r(\bmk) v_{ln}^s(\bmk) v_{nn}^q(\bmk)
    \delta (\varepsilon_{n\bmk} - \varepsilon_{l\bmk} - \hbar \Omega)
    \Bigg]E_r E_s.
\end{align}
If the band structure possesses time-reversal symmetry, which is the case for nonmagnetic materials, then it can shown that the three-velocity term $v_{nl}^r v_{ln}^s v_{nn}^q$ will undergo a sign reversal for $-\bm{k}$: 
$v_{nl}^r(-\bmk) v_{ln}^s(-\bmk) v_{nn}^q(-\bmk) = - v_{ln}^r(\bmk) v_{nl}^s(\bmk) v_{nn}^q(\bmk)$,
In the meantime, the Fermi-Dirac function and the delta function will be even for $\bm{k}$ and $-\bm{k}$. 
As a result, when considering the response to a linearly-polarized light where the $rs$ and $sr$ component cannot be distinguished, the integration of $\bm{k}$ over the Brillouin zone in Eq.~(\ref{eqn:inj}) will be exactly zero, meaning that no contribution will exist for the diagonal part of the density operator. 
(For magnetic systems, $j^{q,\rm{diag}}$ will no longer vanish and is referred to as \textit{injection current}, which will be discussed in more detail in the next subsection.)  

However, $j^{q,\rm{diag}}$ is no zero when additional scattering processes are present. 
To see this, we formally rewrite Eq.~(\ref{eqn:inj}) into a form whose physical meaning is manifest:
\begin{align}
\label{eqn:boltzmann}
    j^{q,\rm{diag}}
    &= 2e\tau_0 \sum_{cv\bmk} \Gamma_{cv,\bmk}^{rs}(\omega) \bqty{ v_{c\bmk}^{q} -
    v_{v\bmk}^{q}},
\end{align}
where we have explicitly considered the transition from the valence band $v$ to conduction band $c$ in a semiconductor due to light excitation. 
The minus sign of $v_{v\bmk}^{q}$ comes from $f_{c\bm{k}} - f_{v\bm{k}} = 0 - 1 = -1$.
$\Gamma_{cv,\bmk}^{rs}(\omega)$ is the carrier generation rate that contains the transition intensity $v_{nl}^r(\bmk) v_{ln}^s(\bmk)$ and the energy selection rule $\delta (\varepsilon_{n\bmk} - \varepsilon_{l\bmk} - \hbar \Omega)$. 
Note we have discretized the integration in the Brillouin zone by a summation over $\bmk$ points.
This is simply the expression for current in the framework of the Boltzmann transport equation, in which the current is equal to the carrier velocity multiplied by its distribution function \cite{Dai21p177403}, and it is expected that without any other interaction, $\Gamma_{cv,-\bmk}^{rs}(\omega)=\Gamma_{cv,\bmk}^{rs}(\omega)$.
If we include additional interactions when computing the carrier generation rate, that is, we extend the Fermi's golden rule to higher orders, then it is likely that $\Gamma_{cv,-\bmk}^{rs}$ is no longer equal to $\Gamma_{cv,\bmk}^{rs}$ in a non-centrosymmetric system, and we call the current from the asymmetric carrier generation \textit{ballistic current}.
One should not confuse ballistic current with \textit{ballistic transport} \cite{Heiblum85p2200} as they carry distinct but related meanings. 
For ballistic transport, it means that the carriers can flow for a certain length without any scatterings, whereas the ballistic current can only exist in the presence of \textit{coherent} scatterings during the optical excitation which induce the population asymmetry (the flow of carriers after the optical excitation will encounter no scatterings for a period $\tau_0$, which is similar to ballistic transport). 

To systematically investigate the effect of interaction on the carrier generation rate, we can express the overall generation rate $\Gamma^{rs}$ in terms of the velocity-velocity (current-current) correlation function $\chi^{rs}$, which essentially counts the number of excited electron/holes by assuming that each absorbed photon will generate an electron-hole pair:
\begin{equation}
\label{eqn:carrier_generation}
    \Gamma^{rs}(\omega) =
    -\frac{2}{\hbar}
    \Im\bqty{\chi^{rs}(\omega)}
    \Big( {\frac{e}{\omega}} \Big)^2 
    E_{r}E_{s},
\end{equation}
where the real-time retarded correlation function $\chi^{rs}$ can be obtained from the corresponding imaginary-time correlation function $\chi_T^{rs}$ via the analytical continuation ${{\chi}^{rs}}(\omega)={{\chi}_{T}^{rs}}({i\omega_n\xrightarrow[]{}\omega+i0^{+}})$, where $0^{+}$ is a infinitesimal positive number \cite{Jishi13Feynman, Mahan13many}. 
In this approach, the interaction effect can be included in the correlation function ${\chi}_{T}^{rs}$:
\begin{align}
\label{eqn:correlation}
    &{\chi_T^{rs}}(i\omega_n) = -\frac{1}{\hbar} \sum_{\bmk\bmk' cc'vv'}
    \braket{v\bmk|\vop^{r}|c\bmk}\braket{c{'}\bmk'|\vop^{s}|v{'}\bmk'} 
    \nonumber \\
    &\times
    \int_{0}^{\hbar/k_{B}T} d{\tau} e^{i{\omega_n}{\tau}}
    \braket{\hat{T}_{\tau} {\crea_{v\bmk}(\tau)} {\anni_{c\bmk}(\tau)} {\crea_{c'\bmk'}(0)} {\anni_{v'\bmk'}(0)} },
\end{align}
from which we can see that the overall carrier generation rate can be decomposed into $\bmk-$resolved carrier generation rate $\Gamma^{rs}(\omega) = \sum_{cv\bmk}\Gamma_{cv,\bmk}^{rs}(\omega)$, and it can be evaluated perturbatively with respect to different interactions. 

Various processes could lead to asymmetric $\Gamma_{cv,\bmk}^{rs}$, such as electron-phonon interaction\cite{Dai21p177403}, electron-hole interaction \cite{Dai21p235203}, and scattering from defects. 
Among these, the electron-phonon interaction and electron-hole interaction are of most interest because they are intrinsic to any semiconductor, regardless of the quality of the crystal. 
So, most work investigating ballistic current will focus on these two interactions.
We would like to have a few more words about these scattering processes being \textit{intrinsic} as some people would instead regard ballistic current as an \textit{extrinsic} mechanism for BPVE. 
Such claim comes from the comparison with shift current where only a perfect and static lattice is considered, so shift current is considered as an intrinsic mechanism, and ballistic current is classified as extrinsic due to the participation of additional processes.
However, this classification will be misleading since any realistic perfect material would have lattice vibration and Coulomb interaction, so we think that ballistic current should also be intrinsic if intrinsic scattering processes are considered.
Nevertheless, different authors could have different philosophies for this classification, and readers should be careful about what they mean by intrinsic and extrinsic.

\begin{figure}
    \centering
    \includegraphics[width=0.45\textwidth]{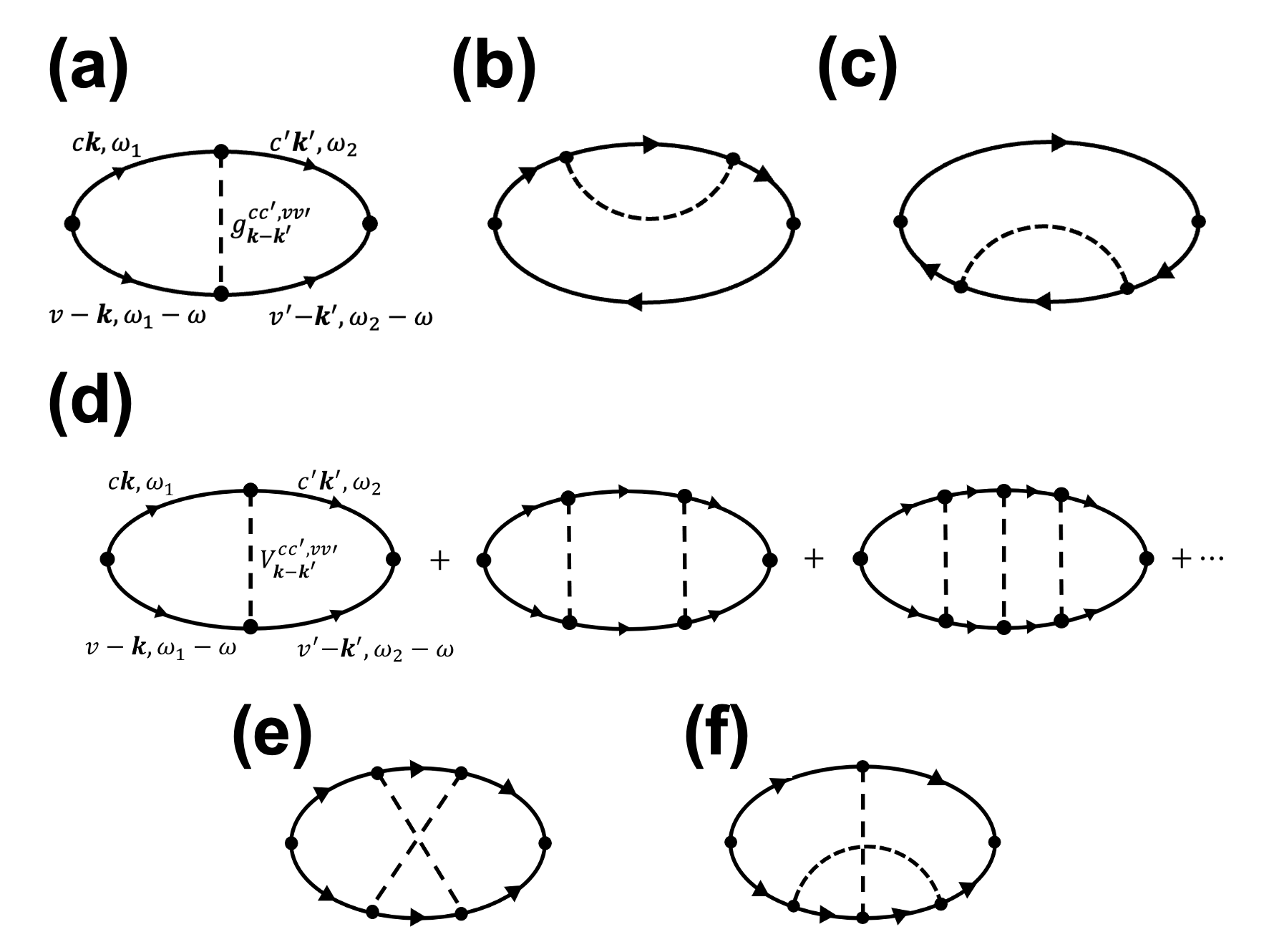}
    \caption{Feynman diagrams for electron-phonon interaction and electron-hole interaction. (a)-(c): Lowest-order diagrams in electron-phonon interaction. Only diagram (a) will give rise to asymmetric scattering. (d)-(f): Diagrams for electron-hole interaction. Unlike the electron-phonon interaction, the electron-hole interaction is known to be long-range, so that higher-order terms cannot be discarded. Nonetheless, it can be shown that only ladder diagrams in (d) will have asymmetric scattering, whereas the diagrams of other types will not.
    Ref.~\cite{Dai21p177403} is reproduced with permission from Phys. Rev. Lett. 126, 177403 (2021). Copyright 2021 American Physical Society.
    Ref.~\cite{Dai21p235203} is reproduced with permission from Phys. Rev. B 104, 235203 (2021). Copyright 2021 American Physical Society.}
    \label{fig:feynman}
\end{figure}

The \textit{ab initio} calculations of ballistic current were realized by Dai, Schankler \textit{et al.}, where the electron-phonon mechanism (termed as phonon-ballistic current) was taken into account \cite{Dai21p177403}. 
By treating the electron-phonon interaction via the Fr\"olich e-ph Hamiltonian, 
\begin{align}
\label{eqn:frolich}
    \hat{H}_{e-\rm{phonon}}
    = \sum_{{\mu}nn{'}} \sum_{\bmk\bmk'} g_{{\mu}\bmk\bmk'}^{nn{'}} 
    {\crea_{n{'}\bmk'}} {\anni_{n\bmk}} {\hat{\Phi}_{\bmk-\bmk'}^{\mu}},
\end{align}
where $\hat{\Phi}_{\tilde{\bm{q}}}^{\mu}=\hat{a}_{\tilde{\bm{q}}}^{\mu}+\hat{a}_{-\tilde{\bm{q}}}^{\mu\dagger}$ is the phonon field operator, $\hat{a}_{\tilde{\bm{q}}}^{\mu}$($\hat{a}_{\tilde{\bm{q}}}^{\mu\dagger}$) are the phonon annihilation(creation) operators, and $g_{{\mu}\bmk\bm{k'}}^{nn{'}}$ is the electron-phonon coupling matrix,
the perturbative expansion of Eq.~(\ref{eqn:correlation}) can be performed with the help of Feynman diagrammatic technique where each perturbative term $\Delta\Gamma_{cv,\bmk}^{rs}(\omega)$ can be represented by a connected diagram. 
For the lowest order non-zero terms, there are three diagrams involved in the optical transition as shown in Fig.~\ref{fig:feynman}(a)-(c), and it can be proved that only Fig.~\ref{fig:feynman}(a) will lead to an asymmetric carrier generation. 
Evaluating this term with Feynman rules, performing the analytical continuation, and using Eq.~\ref{eqn:correlation}, the asymmetric part of the carrier generation rate 
$\Gamma_{cv,\bmk}^{rs, asym}(\omega)
=\frac{1}{2}\Big( \Delta\Gamma_{cv,\bmk}^{rs}(\omega)-\Delta\Gamma_{cv,-\bmk}^{rs}(\omega) \Big)$
can be expressed in terms of velocity matrices $v_{nl}^r(\bmk)$, electron-phonon coupling matrices $g_{{\mu}\bmk\bmk'}^{nn{'}}$, and band energies $\varepsilon_{n\bmk}$. 
The complete form of $\Gamma_{cv,\bmk}^{rs, asym}(\omega)$ can be found in the work \cite{Dai21p177403}, and in combination with Eq.~(\ref{eqn:boltzmann}), the phonon ballistic current can be computed from first-principles calculations.

Similarly, in another work by us \cite{Dai21p235203}, the electron-hole interaction (named as exciton ballistic current) is considered on the same footing as the electron-phonon interaction. 
However, unlike the electron-phonon interaction where it suffices to keep only lowest order terms, the long-range character of the Coulomb interaction will require in principle infinite orders of terms in the perturbative expansion. 
Luckily, most diagrams can be shown not to contribute to the asymmetric scattering, and the sum of infinite orders of ladder diagrams, the ones involved in asymmetric scattering, can be done exactly.  
A certain amount of algebra will lead to rather simple results for the sum of ladder diagrams:
\begin{align}
\label{eqn:correlation_eh}
    &{\chi_{\rm T}^{rs}}(i\omega_n) = -\frac{i}{\hbar} \sum_{\bmk cv}
    \frac{v_{vc}^{r}(\bmk) \tilde{v}^s_{cv}(\omega,\bmk)}
    {\omega+{\varepsilon}_{v\bmk}/\hbar-{\varepsilon}_{c\bmk}/\hbar+i0^{+}},
\end{align}
where
\begin{align}
\label{eqn:effective_v}
    &\mathbf{\tilde{v}}_{cv,\bmk}(\omega) 
    =\braket{c\bmk|\mathbf{\hat{v}}|v\bmk}
    \nonumber \\
    &+\sum_{\bmk' c'v'}{
    \frac{i}{\hbar}
    \frac{V_{\bmk-\bmk'}^{cc',vv'}}
    {\omega+{\epsilon}_{v'\bmk'}/\hbar-{\epsilon}_{c'\bmk'}/\hbar+i0^{+}}
    \mathbf{\tilde{v}}_{c'v',\bmk'}(\omega) 
    }.
\end{align}
Here, $V_{\tilde{\bmq}}^{cc',vv'}$ is the screened Coulomb interaction in the basis of eigenstates of $H_0$, and $\tilde{\bmq}$ is the Fourier component of the Coulomb interaction \cite{Combescot15excitons}. 
Eq.~(\ref{eqn:effective_v}) can be solved numerically to yield $\mathbf{\tilde{v}}_{cv,\bmk}(\omega)$, with which one can calculate the ballistic current from electron-hole interaction from Eq.~(\ref{eqn:inj}), Eq.~(\ref{eqn:carrier_generation}) and Eq.~(\ref{eqn:correlation_eh}).
A different approach to computing exciton ballistic current is also presented in the same work \cite{Dai21p235203} where the Bethe-Salpeter equation is solved, from which the carrier generation rate can be computed from the exciton wave functions.

\begin{figure}
    \centering
    \includegraphics[width=0.4\textwidth]{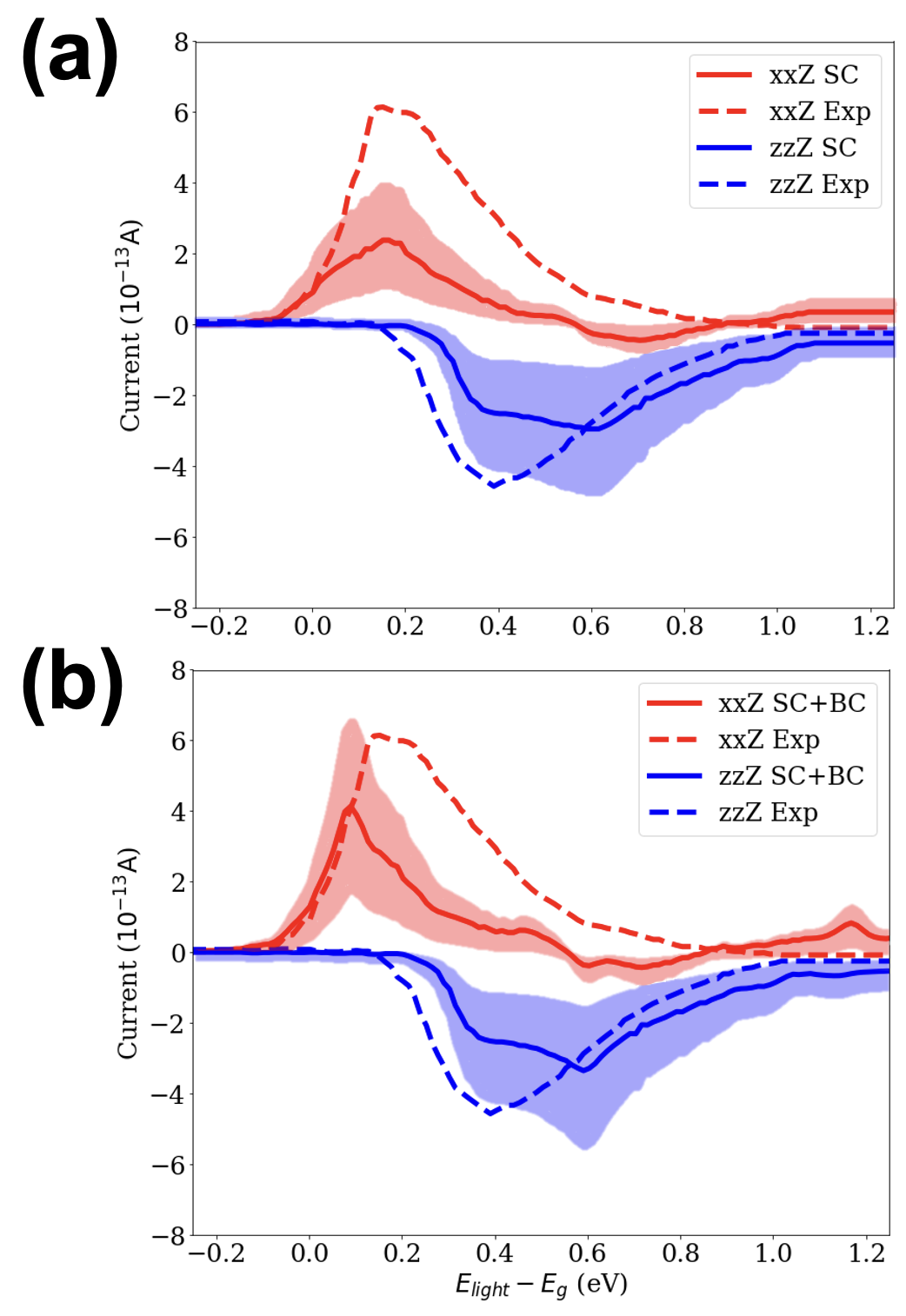}
    \caption{Phonon ballistic current for \ch{BaTiO3}.
    (a) The theoretical shift current has a smaller magnitude than that of the experimental photocurrent.
    (b) After introducing the contribution of ballistic current from electron-phonon coupling, the theoretical BPVE current agrees better with experiments.
    The shaded areas take into account the experimental errors in the sample dimensions, as discussed in Section.~\ref{sec:koch_exp}. 
    Reproduced with permission from Phys. Rev. Lett. 126, 177403 (2021).\cite{Dai21p177403} Copyright 2021 American Physical Society.}
    \label{fig:ph_bc_numercal}
\end{figure}

\begin{figure}
    \centering
    \includegraphics[width=0.48\textwidth]{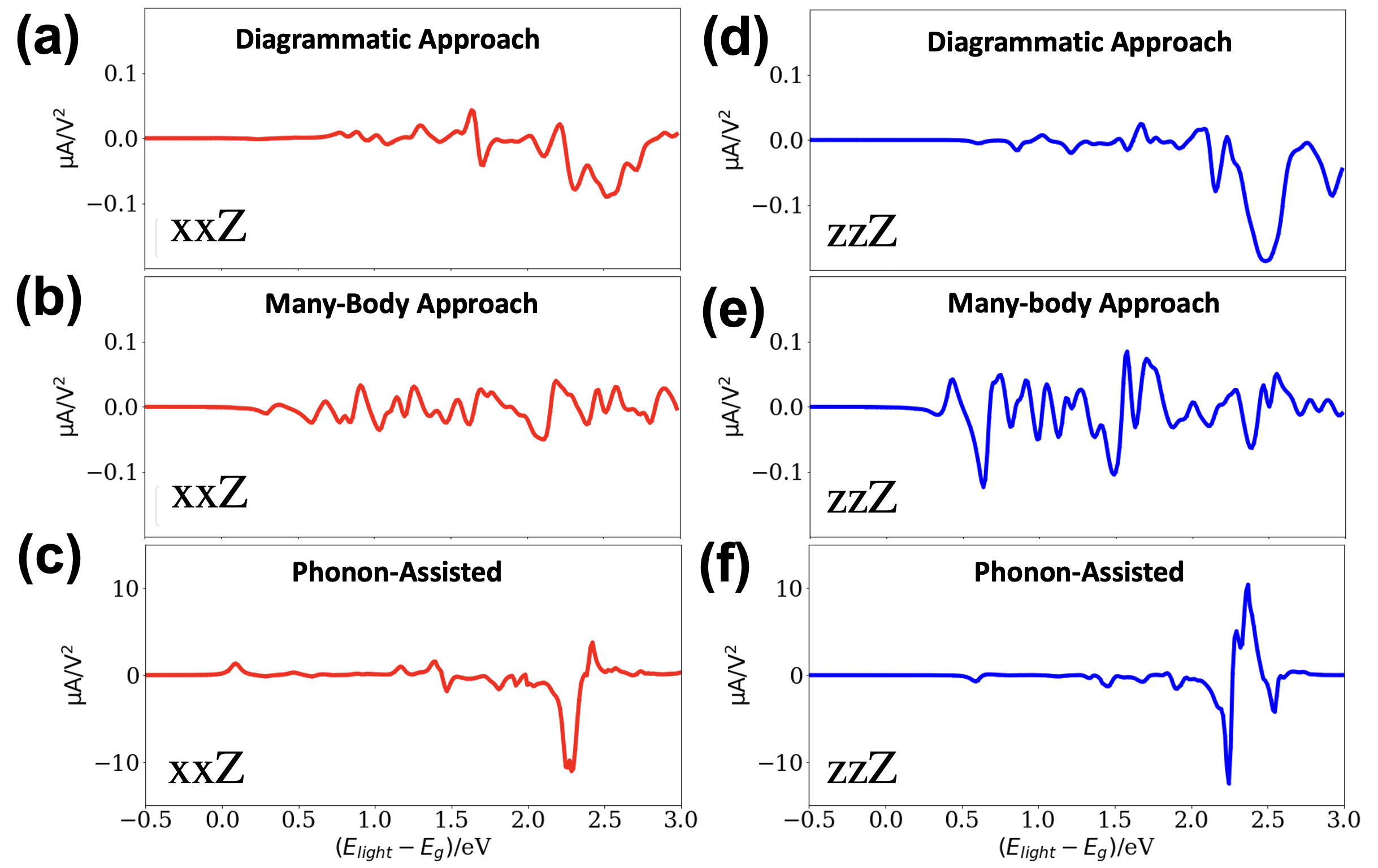}
    \caption{Exciton ballistic current for \ch{BaTiO3}. (a) and (d) Exciton ballistic current from diagrammatic approach. (b) and (e) Exciton ballistic current from the many-body approach. (c) and (f) Phonon ballistic current. Red lines represent the $xxX$ component, and blue lines represent the $xxZ$ component. 
    Clearly, exciton ballistic currents from both methods are much less than the phonon ballistic current.
    Reproduced with permission from Phys. Rev. B 104, 235203 (2021). \cite{ Dai21p235203}Copyright 2021 American Physical Society.}
    \label{fig:ex_bc_numercal}
\end{figure}

According to the procedures prescribed above, the phonon and exciton ballistic currents are calculated for \ch{BaTiO3} and can be found in Fig.~\ref{fig:ph_bc_numercal} and Fig.~\ref{fig:ex_bc_numercal}. 
Note that in Fig.~\ref{fig:ph_bc_numercal} we are plotting the total current (as in Eq.~(\ref{eqn:glass_coeff})) whereas in Fig.~\ref{fig:ex_bc_numercal} we are plotting the response tensor $\sigma_{xx}^z$ and $\sigma_{zz}^z$.
Clearly, the discrepancy between the experimental photocurrent and theoretical shift current for $xxZ$ component can be partially filled by the phonon ballistic current, but for the $zzZ$ component where the shift current is already in good agreement with experiments, the phonon ballistic current barely changes the theoretical photocurrent. 
This shows that in addition to shift current, phonon ballistic current is also a major mechanism in BPVE, and it remains to check whether the exciton ballistic current can further improve the BPVE theory.
Unfortunately, Fig.~\ref{fig:ex_bc_numercal} shows that exciton ballistic current can be two orders smaller than the phonon counterpart, and the smallness makes it hard to connect the features found in the diagrammatic approach with those in the many-body approach. 
Thus, even though we have included infinite orders of Coulomb interaction when computing the asymmetric generation rate (Eq.~(\ref{eqn:carrier_generation})), the canceling among the diagrams makes its overall contribution much smaller than that from the electron-phonon interaction, where only the lowest-order diagram is taken into account.
A similar calculation for monolayer \ch{MoS2} shows similar insights \cite{Dai21p235203}.
To summarize, when evaluating the contribution from ballistic current, it is usually safe to only consider the electron-phonon interaction, and to further improve BPVE theory, scatterings from other sources, such as defects, should be included when computing the asymmetric carrier generation rate.

\subsubsection{Linear Injection Current}
\label{sec:linear_inj}
As alluded above, for nonmagnetic (time-reversal-symmetric, T-symmetry) systems, the injection current will vanish under linearly-polarized light \cite{Lu20p4724, Fei21p207402, Zhang19p3783, Xu21p4330, Wang20p199}. 
As a result, most theoretical and experimental work regarding injection current is centered around the circularly-polarized light, which has a slightly different expression from Eq.~(\ref{eqn:inj}) \cite{Ji19p955, Ni21p154, Ni20p96}.
We will discuss CPGE further in Sec.III.B.
Nonetheless, in recent years, more 2D magnetic materials have been discovered that inspired a renewed interest in their electronic and optical properties, especially in their photovoltaic effect.
As T-symmetry is usually broken in magnetic systems, the symmetry argument about the parity of $v_{nl}^r(\bmk)$ and $\epsilon_{n\bmk}$ no longer applies, which brings about the possibility of observing the injection current even under the linearly-polarized light.

\begin{figure}
    \centering
    \includegraphics[width=0.45\textwidth]{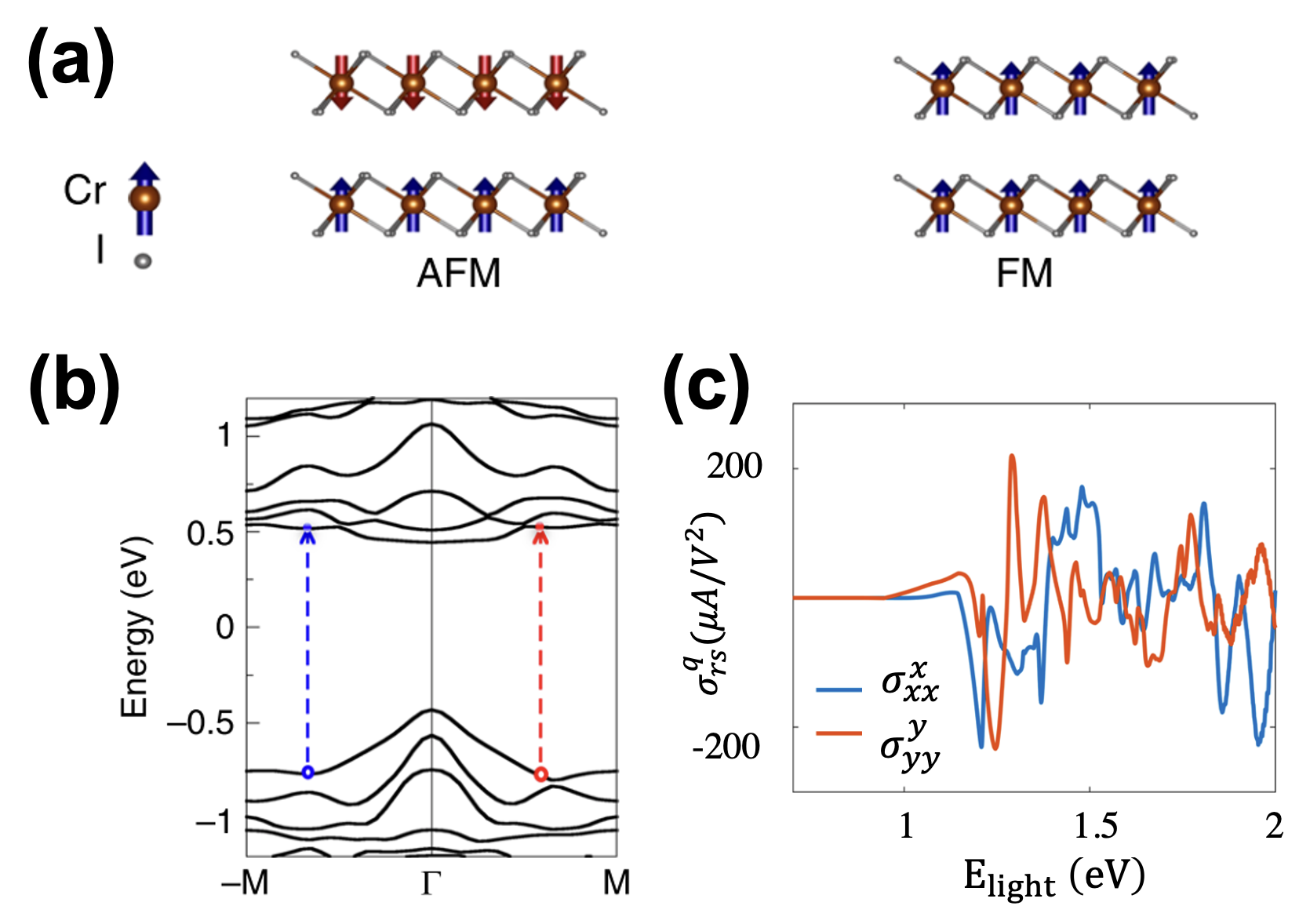}
    \caption{Crystal structure, band structure, and injection current of \ch{CrI3}.
    (a) Spin patterns for AFM and FM phases of bilayer \ch{CrI3}. In the AFM phase, both time-reversal (T) and inversion symmetry (P) are broken, but the combined PT symmetry is preserved. (b) The band structure of the AFM phase \ch{CrI3} in the presence of SOC. 
    The band structure is not symmetric about the $\Gamma$ point, due to the the breaking of T-symmetry combined with SOC.
    (c) The linear injection current for the AFM \ch{CrI3}. When choosing a relatively large relaxation time, for example 600~fs, the linear injection current can be very large and dominates over the shift current (three-band contribution).
    Reproduced with permission from Nat. Commun. 10, 3783 (2019). \cite{Zhang19p3783} Copyright 2019, Springer Nature Limited.}
    \label{fig:cri3}
\end{figure}

One example of the 2D magnetic materials attracting attention is \ch{CrI3}, a ferromagnetic insulator \cite{Zhang19p3783, Fei21p207402}. 
In the bilayer case, it can exhibit two phases: ferromagnetic (FM) and antiferromagnetic (AFM) (Fig.~\ref{fig:cri3}). 
The latter will break both the inversion symmetry and T-symmetry, causing an asymmetric band structure at $\bmk$ and $-\bmk$.
Furthermore, the band velocities at $\bmk$ and $-\bmk$ will not cancel so that there is no requirement that the carrier generation rates at opposite $\bmk$ points will be equal. 
Therefore, by using Eq.~(\ref{eqn:inj}), the injection current of \ch{CrI3} can be calculated (Fig.~\ref{fig:cri3}(c)).
A similar investigation of \ch{MnBi2Te4} has also been done \cite{Wang20p199}, demonstrating a giant injection current (two orders higher than the shift current of \ch{PbTiO3} and \ch{BaTiO3}.)

Two things to note about these calculations: 1. The AFM phase of \ch{CrI3} is special in that its centrosymmetry is broken by spins.
So, an inversion operation about the interlayer inversion center will keep the lattice the same but reverse the spin directions.
Thus, the spin-orbital coupling (SOC) is required to make sure that the AFM and reverse-AFM will correspond to different energies.
Otherwise, neglecting the SOC will make the band structure still symmetric for $\bmk$ and $-\bmk$ \cite{Zhang19p3783, Fei21p207402}.
2. The relaxation times $\tau_0$ used in these works \cite{Wang20p199, Zhang19p3783} are obtained from experimental values of materials belonging to the same family and are somewhat arbitrary. 
So, the large injection current observed in these calculations is partly due to the large relaxation time. 
A more consistent treatment for relaxation time would be calculating $\tau_0$ from first principles as is done by Dai, Schankler $\textit{et al.}$ \cite{Dai21p177403} when computing the ballistic current. 
Their calculations show that the constant relaxation time approximation is reasonable, showing weak dependence on band indices and crystal momenta, but the value would differ from material to material.
For example, the computed momentum relaxation time of \ch{BaTiO3} is 2~fs, compared with 100~fs and 600~fs used in \ch{MnBi2Te4} and \ch{CrI3}, respectively.
Thus, when interpreting the magnitude of injection current calculated with the constant relaxation time approximation, one has to be careful about the choice of the $\tau_0$, and this consideration also holds for circular injection current (Sec.~\ref{sec:theory_circular_light}).

Another scenario where the linear injection current could be important is the photo-generation of pure spin current (PSC) \cite{Fei21arxiv, Wang20p199}.
In this case, T-symmetry-breaking is not needed because there is no charge current.
What is required for linear injection PSC, however, is a sizable SOC which will make the band structure at $\bmk$ and $-\bmk$ have different spin characters, as shown in Fig.~\ref{fig:psc}.
If we use Eq.~(\ref{eqn:inj}) to compute the spin-up and spin-down injection currents separately, the band structure of each spin is asymmetric even though the overall band structure is symmetric.
Then, spin-up carriers will have a net current whose direction is opposite to spin-down carriers, generating a pure spin current. 
One can immediately see why strong SOC is a prerequisite for injection PSC as for systems with no or weak SOC, the spin-polarized band structure will become symmetric again.
Therefore, it is expected to observe large PSC in materials having heavy elements, such as \ch{CdS}, \ch{SnTe}, and transition metal dichalcogenides (TMD).
On the hand, in terms of generating PSC, shift current is also predicted to be a viable mechanism \cite{Young13p057201}.
In contrast to the linear injection current, the shift current mechanism does not necessarily require SOC.
Instead, it can exist in antiferromagnetic systems.
Take the PT-symmetric (breaking P- and T-symmetry individually but preserving the PT as a whole) hematite \ch{Fe2O3} for example, the PT-symmetry will make shift vectors at the same $\bmk$ point have opposite directions for spin-up and spin-down components despite the same transition rate, thus giving rise to a pure spin current.
The general symmetry requirements for using shift current to generate PSC has been identified in the work\cite{Young13p057201}, and its existence is demonstrated by performing first-principles calculations on several antiferromagnetic materials, such as \ch{Fe2O3} and \ch{BiFeO3}. 

\subsubsection{Kinetic Model}
Having discussed several major mechanisms for BPVE under linearly-polarized light, it could be inspiring to unify them in the perspective of a kinetic model.
A kinetic model takes into account various processes (excitation, recombination, etc.) contributing to the time-evolution of the density matrix, including the diagonal (occupation) and off-diagonal (coherence) matrix elements.
In principle, it is able to describe temporal, steady-state, and equilibrium time evolution, while most experiments measure observables in steady-state or equilibrium states, in which the density matrix elements should possess stable values and can thus be used to compute these observables. 
Therefore, to study the steady-state DC current due to BPVE, it is desirable to find the influence of all relevant photoinduced and thermal processes and connect them in a quantum Liouville equation in order to establish the steady-state values in a kinetic model.

This kinetic model was originally conceived by Belinicher {\em et al.}, where several important processes were considered, including light excitation, electron-phonon coupling, and defect scattering \cite{Belinicher82p649}.
As pointed out by Belinicher {\em et al.}, the foundational idea of the kinetic model is that the time-evolution of the density operator can be described by the quantum Liouville equation (written in the Schr\"{o}dinger picture):
\begin{align}
    \label{eqn:liouville_general}
    &i\hbar \dot{\oprho}(t) = [\hat{H}, \oprho(t)], 
    \nonumber \\
    &\hat{H} = \hat{H}_0 + \hat{H}_{e-\mathrm{light}} 
    + \hat{H}_{e-\mathrm{phonon}}
    + \hat{H}_{\rm Coulomb}
    + \hat{H}_{e-\mathrm{defect}},
\end{align}
where $H_{e-\mathrm{light}}$ is quantized (the vector potential is expressed as photon creation and annihilation operators) and drives the excitation (including stimulated recombination if $T>0$) and spontaneous recombination, while $\hat{H}_{e-\mathrm{phonon}}$ and $\hat{H}_{e-\mathrm{defect}}$ are responsible for the thermalization in the full kinetic cycle, and they can also participate in the excitation and recombination processes as well in the full kinetic cycle.
The quantum treatment of light can enable the spontaneous recombination because it will allow the electronic system to be coupled to the light that is at every possible frequency, so there is a driving force for electron and hole to recombine even though their energy difference is different from that of the incident light.
The electrical current calculated from $j=e/m \Tr[\oprho \hat{j}]$ can thus be categorized as excitation, thermalization, and recombination current according to the processes participating in the current generation.

The progress reported in this review can be described as developing first-principles computational approaches that provide some terms in the kinetic model. 
The rest of these terms can be approximated, with sensible functional forms.
For example, in deriving the shift current and ballistic current, we are essentially computing the current generated in the excitation process, so for this purpose, we take $\hat{H}_{e-\mathrm{light}}$ as composed of only monochromatic light (which is equivalent to taking it as a classical monochromatic electromagnetic field), and we approximate the thermalization and spontaneous recombination process related to $\hat{H}_{e-\mathrm{phonon}}$ and $\hat{H}_{e-\mathrm{defect}}$ by the constant-relaxation-time approximation:
\begin{align}
    \label{eqn:liouville}
    i\hbar \dot{\oprho}(t) = [\hat{H}_0 + \hat{H}_{e-\mathrm{light}}, \oprho(t)] 
    - i \hbar \frac{\oprho(t) - \oprho_0}{\tau_0},
\end{align}
with $\hbar / \tau_0 = \eta$.
The last term concerns the dissipation that takes into account the thermalization, which would otherwise be taken care of by $\hat{H}_{e-\mathrm{phonon}}$ and $\hat{H}_{e-\mathrm{defect}}$, and the recombination (the carriers no longer recombine spontaneously through $\hat{H}_{e-\mathrm{light}}$ since it now represents a classical field and cannot absorbed the emitted photons).
To compute shift current, we consider how the off-diagonal elements of $\oprho(t)$ evolve according to Eq.~(\ref{eqn:liouville}), which will lead to Eq.~(\ref{eqn:brute_force_j}) and Eq.~(\ref{eqn:shift_current}).
On the other hand, if we are interested in computing the ballistic current, which is from the diagonal part of $\oprho(t)$, then we need to include $\hat{H}_{e-\mathrm{phonon}}$ and $\hat{H}_{\mathrm{Coulomb}}$ in Eq.~(\ref{eqn:liouville}) and consider their contributions to the diagonal elements of  $\oprho(t)$ \textit{only at} the excitation process. 
To be more specific, at the moment of the optical excitation, the scatterings from $\hat{H}_{e-\mathrm{phonon}}$ and $\hat{H}_{\mathrm{Coulomb}}$ will interfere with the scatterings from $\hat{H}_{e-\mathrm{light}}$ and give rise to the phonon ballistic current, while the thermalization and recombination processes
\footnote{As a side note, one may find that $\eta$ also has the same functional form as in the classic formulation of ``adiabatic turning-on'' used when calculating the absorption in the linear response theory.
In this approach, the periodic perturbation is multiplied by $e^{\eta t}$ to break the perfect periodicity \cite{Giuliani05quantum}. 
The presence of $\eta$ broadens the resonance from a single frequency to a range of frequencies, and as a result, the system can reach steady-state and keep constant energy instead of growing in energy as in a perfect resonance process.
Physically speaking, the absorbed energy will be dissipated to the environment via the implicit coupling characterized by $\eta$.
In this sense, the $\eta$ appearing in the adiabatic turning-on is performing the same role as the $\eta$ in Eq.~(\ref{eqn:liouville}).} 
are still approximated by $i \hbar(\oprho(t) - \oprho_0)/{\tau_0}$.

Some additional types of bulk photovoltaic current originating from the more general expression Eq.~(\ref{eqn:liouville_general}) have been formulated already \cite{Belinicher82p649}, while their reformulation into first-principles calculations are still ongoing.
It has been shown by Belinicher {\em et al.} that in addition to the excitation shift current, there also exist real-space shift currents associated with the thermalization and recombination.
The recombination shift current is easily evaluated, since its form should be exactly the same as the excitation shift current Eq.~(\ref{eqn:shift_current}) except that the distribution functions will be replaced by the non-equilibrium steady-state distribution.
Hence, it is mostly concentrated at the band edge states with sign opposite to the excitation shift current.
The thermalization shift current, or phonon shift current, accompanies the electron-phonon scattering processes \textit{after} the optical excitation.
The related shift vector is similar to Eq.~(\ref{eqn:sv}) with the phase being changed to the phase of the electron-phonon coupling matrix elements \cite{Belinicher82p649}.
This contribution could be important, since the phonon ballistic current plays an important role in \ch{BaTiO3} \cite{Dai21p177403, Dai21p235203}. Other possibilities also exist, for example in the spontaneous recombination process if electron-phonon and electron-hole interactions are taken into account, so there are still numerous opportunities in this regard. Being able to evaluate the kinetic model from first principles will greatly enhance the accuracy and the predictive power of the BPVE theory, and this kinetic model devised for steady-state could be potentially extended to compute the temporal evolution of the density matrix in order to study ultrafast experiments.

\subsubsection{Relation to Anomalous Hall Effect}
At this point, readers who are familiar with the anomalous Hall effect (AHE) may have noticed that the shift current and ballistic current in BPVE have direct parallels with the \textit{side jump} and \textit{skew scattering} mechanisms in AHE \cite{Nagaosa10p1539}. 
It is interesting however that no comparison has been made for the two phenomena; in this section, we make the connection explicit.
AHE in itself is a vast topic and includes many different aspects for its theory development, so this section is by no means comprehensive.
Readers are encouraged to read the review by Nagaosa \textit{et al.} \cite{Nagaosa10p1539} to learn more about AHE.

AHE is the phenomenon that when measuring the Hall effect in a ferromagnetic metal, the Hall current deviates from the Lorentz law and is usually very large.
From this description, it is clear that AHE usually refers to the linear transverse response to static electric field, and for it to happen, breaking T-symmetry is required. 
For BPVE, however, it is the second-order response to the oscillating electric field (light), and it is not restricted to the current generation in a direction transverse to the electric field of light.
Rather than breaking T-symmetry, breaking inversion symmetry is required for BPVE, which is is due to the characteristics of second-order response as discussed in the beginning of the Section.~\ref{sec:theory_linear_light}.
So, it can certainly be seen that AHE and BPVE describe two distinct phenomena.

However, when formulating the theories for AHE and BPVE, one can often find that the ideas developed in AHE can be borrowed to understand BPVE.
There are two extrinsic mechanisms induced by defects in AHE, namely skew-scattering and side-jump.
It is generally accepted that skew-scattering means the breaking of detailed balance (transition rate of $\bmk \rightarrow -\bmk$ no longer equals to that of $-\bmk \rightarrow \bmk$) in the presence of scattering processes (mostly scattering from magnetic impurities or disorders) with strong SOC and breaking of T-symmetry, which makes that the carrier generation rate have a preferred direction and thus asymmetric. 
This is very similar to the idea rooted in the ballistic current for BPVE, but in BPVE the strong SOC and breaking of T-symmetry are not required, and scattering processes usually refer to electron-phonon scattering and electron-hole scattering.

The side-jump mechanism describes the coordinate shift of the electron wave packet when scattered by a magnetic impurity with SOC, which resembles the shift current mechanism for BPVE.
Indeed, the expression for coordinate shift in side-jump mechanism is exactly same as the shift vector in shift current mechanism except that the transition rate is now governed by the impurity scattering in AHE instead of the optical excitation in BPVE \cite{Sinitsyn06p075318}.
One can try to replace the transition rate in side-jump with the optical transition rate (transition dipole matrix) and then recover the shift current expression in a semiclassical treatment of AHE.
Thus, it is expected that the further development of BPVE theory could continue to be inspired by the better-understood AHE phenomenon.
Conversely, a novel nonlinear AHE effect due to the so-called Berry curvature dipole is predicted in non-magnetic but non-centrosymmetric systems, which clearly has its inspiration from the theory of BPVE \cite{Sodemann15p216806}.

\begin{figure}
    \centering
    \includegraphics[width=0.45\textwidth]{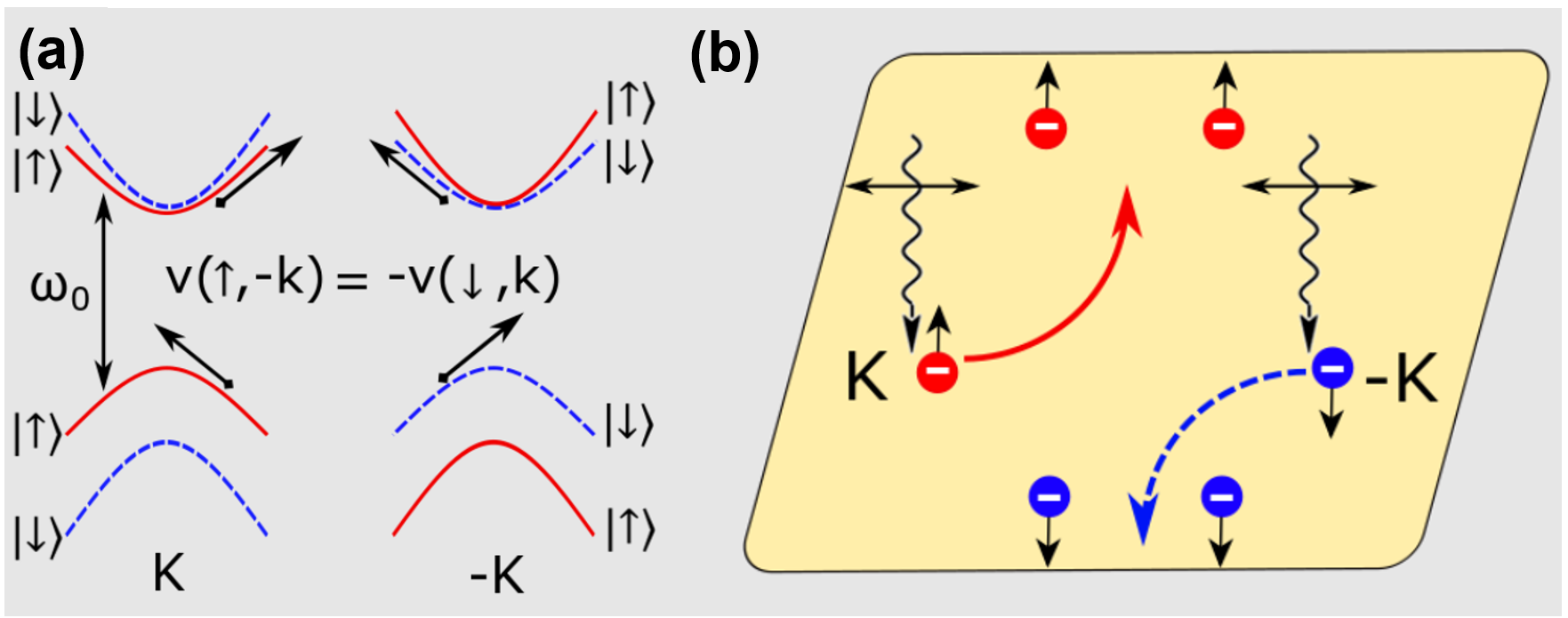}
    \caption{Pure spin current (PSC) from injection current under linearly-polarized light in a non-magnetic system.
    Due to the existence of SOC, the bands at opposite $\bmk$ points in the Brillouin zone would have different spin indices. As a result, under the optical excitation from linearly-polarized light, electrons at K will move in a direction opposite to the electrons at -K, leading to a zero net charge current. But because of their different spin direction, a spin current will be generated along with zero charge current, causing a pure spin current.
    Reproduced from the work \cite{Fei21arxiv}.
    }
    \label{fig:psc}
\end{figure}

\subsection{\label{sec:theory_circular_light} Circularly Polarized Light}
After finishing the discussion of the basic aspects of the BPVE under linearly-polarized light, it is natural to examine the photo-response under the circularly-polarized light. 
This phenomenon has a more well-known name in the area of spectroscopy, the circular photogalvanic effect (CPGE) \cite{Ji19p955, Ni20p96, Ni21p154}.
So, in this section, we refer to the circular BPVE as CPGE to be consistent with the existing literature.
Different from the linear BPVE Eq.~\ref{eqn:phenom}, the phenomenological description of CPGE can be written as:
\begin{align}
    \label{eqn:phenom_cir}
    j^q = i\gamma_{ql}[\bm{E} \times \mathbf{E^*}]_l.
\end{align}
Here, $\gamma_{ql}$ is the response tensor for CPGE, $q$ is again the Cartesian direction of the photocurrent, and $l$ is the propagation direction of the circularly-polarized light.
Note that $\bm{E}$ is vector form of the electric field, so for circularly-polarized light, it has the form $\bm{E}=(E_x,~e^{\pm i\frac{\pi}{2}} E_y,~0)$ with $E_x = E_y = E_0$ (assuming the light is propagating along $z$, and +(-) representing left(right) helicity). 
Then, $i[\bm{E} \times \mathbf{E^*}]_l = \pm (E_r E_s - E_s E_r)$, so the second-order response must differentiate between $E_r E_s$ and $E_s E_r$ in order to have non-zero response.

Now looking at the ballistic current Eq.~(\ref{eqn:boltzmann}) and the more explicit expressions for the carrier generation rate \cite{Dai21p177403}, one can find that there is no requirement that the $rs$ component has to be equal to $sr$, so the ballistic current from electron-phonon interaction and electron-hole interaction can also appear for circularly-polarized light. 
More interestingly, the intrinsic diagonal contribution (in the absence of extra scattering processes) will be non-zero as well for the same reason, and this non-vanishing injection current is actually what people usually refer to as the CPGE.
For shift current, however, the derivation of Eq.~(\ref{eqn:shift_current}) has already symmetrized the $rs$ and $sr$ components due to the consideration that linearly-polarized light cannot distinguish $rs$ and $sr$.
Therefore, Eq.~(\ref{eqn:shift_current}) cannot be used directly to analyze the shift current under circularly-polarized light, and different expressions need to be worked out in this case. 
Below, we examine the injection current and shift current more closely for left-handed circularly-polarized light, and the results for the right helicity can be obtained with a sign reversal.

\subsubsection{Circular Shift Current}
The derivation of shift current under circularly-polarized light is largely the same as that under linearly-polarized light, except that now different components of the vector field will have different phases by $\pi/2$: $\bmaop^{CP}(t) = \hat{A}_r \bm{e_r} \cos{\omega t} + \hat{A}_s \bm{e_s} \sin{\omega t}$, where $\bm{e_r}$ and $\bm{e_s}$ are the unit vectors representing the polarization directions.
Then, the perturbation will become:
\begin{align}
    \label{eqn:cir_e-light}
    \hat{H}^{CP}_{e-\rm{light}} &= \frac{e}{m}\bmvop \cdot \bmaop^{CP}(t) 
    \nonumber \\
    &= \frac{e}{m}\bmvop \cdot (\hat{A}_r \bm{e_r} \cos{\omega t} + \hat{A}_s \bm{e_s} \sin{\omega t}), 
\end{align}
and the density operator Eq.~(\ref{eqn:pertur_rho}) will be expanded under the new perturbation, and the off-diagonal contribution (responsible for shift current) can be separated by requiring $n \neq m$ in Eq.~(\ref{eqn:current_density_op}).
Following exactly the same procedure as for linear shift current, we can arrive at an important expression for the off-diagonal contribution \cite{Gao21p042032}:
\begin{align}
    \label{eqn:cp_off}
    &j^{q,CP-sh} = E_r E_s 
    \frac{e^3}{\omega^2} \sum_{n, l} \sum_{\Omega=\pm \omega}\int d\bmk 
    \left(
    f_{n\bmk} - f_{l\bmk}
    \right)
    \nonumber \\
    &\times
    \bigg[
    i v^s_{nl}(\bmk) \frac{\partial v^r_{ln}(\bmk)}{\partial k_q}
    -i v_r^{nl}(\bmk) \frac{\partial v^s_{ln}(\bmk)}{\partial k_q}
    \nonumber \\
    &-\left(
    \chi_{lq}(\bmk) - \chi_{nq}(\bmk)
    \right)
    \left(
    v^s_{nl}(\bmk) v^r_{ln}(\bmk) - v^r_{nl}(\bmk) v^s_{ln}(\bmk)
    \right)
    \bigg]
    \nonumber \\
    &\times 
    \left(
    i \pi \delta(\varepsilon_{n\bmk} - \varepsilon_{l\bmk} + \hbar \Omega)
    + \pv{\frac{1}{\varepsilon_{n\bmk} - \varepsilon_{l\bmk} + \hbar \Omega}}
    \right),
\end{align}
where $\pv$ represents the principal part integration and all the other symbols carry the same meaning as those in Eq.~(\ref{eqn:shift_current}).

With T-symmetry, the terms multiplied by the delta function $\delta(\varepsilon_{n\bmk} - \varepsilon_{l\bmk} + \hbar \Omega)$ can be shown to vanish by considering the parities at $\bmk$ and $-\bmk$ as well as by switching the dummy indices $n$, $l$, and $\Omega$.
However, the terms mulitplied by the principal part $\pv{\frac{1}{\varepsilon_{n\bmk} - \varepsilon_{l\bmk} + \hbar \Omega}}$ will survive, giving rise to a mysterious non-resonant (sub-bandgap) response, which we will denote as $j_{pp}^{q,CP-sh}$. 
One is therefore attempted to conclude that there exists sub-bandgap shift current for circularly-polarized light.
However, there is no concrete experimental result the showing existence of such a non-resonant shift current, so this term is apparently unphysical and has to be canceled by some arguments or by other unknown terms. 
Fortunately, we can indeed identify such a term in the diagonal contribution of density operator $\oprho^I$, which makes the shift current vanish under circularly-polarized light for systems possessing T-symmetry, and it is the combination of $j_{pp}^{q,CP-sh}$ with this term from $\rho_{nn}^I$ that is responsible for a distinct additional contribution in metals (See Section.~\ref{sec:fermi_contribution}).

\subsubsection{Circular Injection Current}
The diagonal contribution $\rho_{nn}^I$ in Eq.~(\ref{eqn:current_density_op}) can be derived in a similar fashion as the off-diagonal contribution, but under circularly-polarized light, it is worth pausing at an intermediate step and examining the derivation:
\begin{align}
    \label{eqn:cir_diag_interm}
    &j^{q,CP-\rm{diag}} = E_r E_s 
    \frac{e^3}{2 \omega^2} \sum_{n, l} \sum_{u=\pm 1} \int d\bmk 
    \left(
    f_{n\bmk} - f_{l\bmk}
    \right) 
    v_q^{nn}(\bmk)
    \nonumber \\
    &\times
    \frac{1}{ \eta (\varepsilon_{n\bmk} - \varepsilon_{l\bmk} + \hbar \Omega - i\eta)}
    \left(
    v^s_{nl}(\bmk) v^r_{ln}(\bmk)
    -v_r^{nl}(\bmk) v^s_{ln}(\bmk)
    \right).
\end{align}
Apparently, when the infinitesimal $\eta \rightarrow 0$, Eq.~(\ref{eqn:cir_diag_interm}) will diverge (ballistic current will also diverge if there is no relaxation, $\tau_0 \rightarrow \infty$).
But in the constant relaxation-time approximation $\eta=\hbar/\tau_0$ where $\tau_0$ is the relaxation time, as long as the excited carriers are scattered (which is always the case in real materials), the $\tau_0$ will be a finite value and thus the injection current will not diverge.
On the other hand, even in the clean limit where $\tau_0 \rightarrow \infty$, the \textit{injection rate} $d j^{q,CP-\rm{diag}}/d\tau_0$ will remain finite and constant; this is why people name this current ``injection current'' in the first place because it seems to indicate that the light is injecting carriers into the system with a constant rate regardless of the relaxation time (this is of course only valid when $\eta$ is still reasonably small).
So, the usual interpretation or definition for injection current is:
\begin{align}
    \label{eqn:def_cp_inj}
    j^{q,CP-\rm{inj}}=\tau_0 (d j^{q,CP-\rm{diag}}/d\tau_0 ),
\end{align}
which has an intrinsic term, injection rate, that does not depend on the scattering mechanisms of materials, and an extrinsic term, the relaxation time, that varies with relaxation mechanisms \cite{Gao21p042032, Sipe00p5337, Parker19p045121}. 

\begin{figure}
    \centering
    \includegraphics[width=0.48\textwidth]{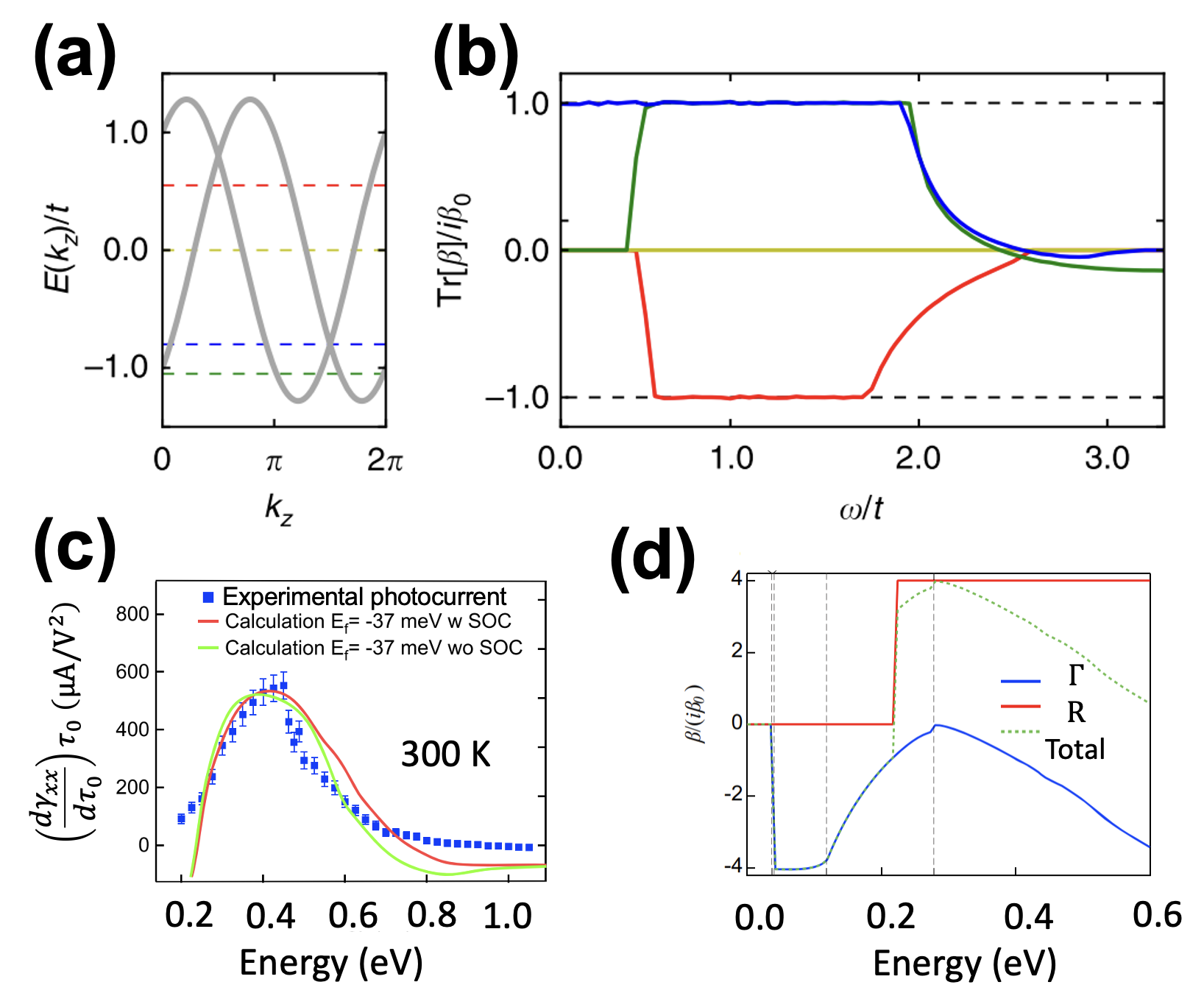}
    \caption{Quantized CPGE.
    (a) Band structure for a generic two-band Weyl semimetal model. (b) CPGE trace for the same model, for four different values of the chemical potential ($\mu/t$ = 1.05, -0.8, 0.0, 0.55) represented as dashed lines in (a). For frequencies between the Weyl node energies the CPGE trace is quantized. 
    (c) Experimentally measured CPGE for a Weyl-semimetal \ch{CoSi} and the theoretically calculated CPGE with different parameters. The experimental CPGE shows a plateau feature which can be taken as quantized response.
    (d) The $R$ and $\Gamma$ point contributions to the overall CPGE computed from a $\bmk \cdot \mathbf{p}$ model, which can show where the plateau (quantized CPGE) is coming from.
    Ref.~\cite{Juan17p15995} is reproduced with permission from Nat. Commun. 8, 15995 (2017). Copyright 2017 Springer Nature Limited.
    Ref.~\cite{Ni21p154} is reproduced with permission from Nat. Commun. 12, 154 (2021). Copyright 2021 Springer Nature Limited.}
    \label{fig:quantized_cpge}
\end{figure}

Proceeding with the definition Eq.~(\ref{eqn:def_cp_inj}), we can get a well-known expression for circular injection current or CPGE for semiconductors:
\begin{align}
    \label{eqn:cir_inj}
    j^{q,CP-\rm{inj}} = & iE_r E_s 
    \frac{\pi e^3  \tau_0}{\hbar \omega^2} \sum_{v, c} \int d\bmk 
    \left(v^q_{cc}(\bmk) - v^q_{vv}(\bmk) \right)
    \nonumber \\
    &\times
    \left(v^r_{cv}(\bmk) v^s_{vc}(\bmk)
    -v^r_{vc}(\bmk) v^s_{cv}(\bmk)
    \right)
    \delta(\varepsilon_{v\bmk} - \varepsilon_{c\bmk} + \hbar \omega ),
\end{align}
and a slightly-modified expression for metal can be easily obtained by considering the partial occupation of the Fermi-Dirac function.
Like shift current, the circular injection current is also related to the phase of wave functions from the $\Im{v^r_{cv}(\bmk) v^s_{vc}(\bmk)}$ and is reminiscent of the Kubo formula for Hall conductivity. 
Indeed, it has been shown that the trace of the circular injection current response tensor can be quantized for a two-band model of Weyl semimetal\cite{Juan17p15995}, as shown in Fig.~\ref{fig:quantized_cpge} (we always assume perfect linear dispersion, i.e. within the Lifshitz energy).  
The exact quantization makes use of an idealized Weyl node (which is perfectly linear with $\bmk$) and the fact that only two bands are involved in the transition, which would allow us to rewrite the trace of $d\gamma_{ql} / d\tau_0$ extracted from Eq.~(\ref{eqn:cir_inj}) (comparing with Eq.~(\ref{eqn:phenom_cir})) in a form
that is dependent of the monopole charge $C_L$ of the Weyl node:
\begin{align}
    \label{eqn:quantized_cpge}
    \Tr[\frac{d\gamma}{d\tau_0}] = i\pi \frac{e^3}{h^2} C_L.
\end{align}
Later, an experiment measuring the CPGE on a Weyl semimetal \ch{CoSi} confirms this quantization within experimental resolution \cite{Ni21p154}. 
To be specific, at certain photon energies, only some regions adjacent to Weyl points in the Brillouin zone are responsible for the observed CPGE, and the optical transitions occur between the two bands composing the Weyl nodes. 
Accordingly, the circular injection current is seen to have dips and plateaus (Fig.~\ref{fig:quantized_cpge}d) that agree with the predicted quantization.

There are other variations of Eq.~(\ref{eqn:cir_inj}) that are used to explain or predict novel phenomena. 
In the work by Ji \textit{el al.}\cite{Ji19p955}, the spatial inhomogeneity of the light spot is taken into account to rationalize the sign flip of photocurrent when measured at different positions of the sample.
In another work by Fei \textit{et al.}\cite{Fei21p207402}, Eq.~(\ref{eqn:cir_inj}) is adapted so that the spin freedom is shown explicitly, and then they demonstrate that the circular injection current can be used as a mechanism to generate spin current in PT-symmetric antiferromagnetic insulators (breaking P- and T-symmetry individually but preserving the PT-symmetry). 
The basic idea is that the circularly-polarized light will make the $\Im{v^r_{cv}(\bmk) v^s_{vc}(\bmk)}$ have different signs for spin-up and spin-down electrons under PT-symmetry, causing their respective photocurrents to flow along opposite directions.
In contrast to the PSC generated for the nonmagnetic materials using linear injection current, which requires a large SOC (Fig.~\ref{fig:psc}) \cite{Fei21arxiv}, the spin current generated via circular injection current mechanism is insensitive to SOC and therefore fundamentally different from the linear scenario, though it is quite similar to the linear shift current mechanism in generating PSC.
A summary of using BPVE for generating charge and spin current can be found in the work \cite{Xu21p4330}.

To close the discussion on circular injection current, we want to emphasize that some important information will be lost when using the interpretation of injection rate $j^{q,CP-\rm{diag}}/d\tau_0$, as pointed out by Gao \cite{Gao21p042032}. 
To see this, let's rewrite the energy term in Eq.~(\ref{eqn:cir_diag_interm}) as:
\begin{align}
    \label{eqn:cir_diag_ene_term}
    &\frac{1}{ \eta (\varepsilon_{n\bmk} - \varepsilon_{l\bmk} + \hbar \Omega - i\eta)}
    \nonumber \\
    &=\frac{\varepsilon_{n\bmk} - \varepsilon_{l\bmk} + \hbar \Omega + \eta}
    { \eta [(\varepsilon_{n\bmk} - \varepsilon_{l\bmk} + \hbar \Omega)^2 + \eta^2 ]} 
    +\frac{i}
    { \eta [(\varepsilon_{n\bmk} - \varepsilon_{l\bmk} + \hbar \Omega)^2 + \eta^2] }.
\end{align}
The second term at the RHS of Eq.~({\ref{eqn:cir_diag_ene_term}}) is weakly dependent on $\tau_0$ and will thus be dropped when taking the derivative $j^{q,CP-\rm{diag}}/d\tau_0$, but it turns out that the current associated with this term, which we denote as $j_{pp}^{q,CP-\rm{diag}}$, is totally physical and can cancel the sub-bandgap contribution from $j_{pp}^{q,CP-sh}$.

\subsubsection{Fermi Surface Response}
\label{sec:fermi_contribution}
To see how $j_{pp}^{q,CP-\rm{diag}}$ and $j_{pp}^{q,CP-sh}$ combine, the first step is to realize that by symmetrizing $n$ and $l$ in Eq.~(\ref{eqn:cp_off}) and Eq.~(\ref{eqn:cir_diag_interm}), the terms involving the derivative of velocity matrices in $j_{pp}^{q,CP-sh}$ can be rewritten in the form of:
\begin{align}
    \label{eqn:v_derive_rewrite}
    &v^s_{nl} \frac{\partial v^r_{ln}}{\partial k_q}
    -v_r^{nl} \frac{\partial v^s_{ln}}{\partial k_q}
    -\frac{\partial v_r^{nl}}{\partial k_q} v^s_{ln} 
    +\frac{\partial v^s_{nl}}{\partial k_q} v^r_{ln} 
    \nonumber \\
    =&\frac{\partial}{\partial k_q}
    \left(
    v^s_{nl} v^r_{ln}
    -
    v_r^{nl} v^s_{ln}
    \right),
\end{align}
where the the $\bmk$-dependence has become implicit for conciseness, and the Berry connection terms can be shown to vanish by switching $n$ and $l$.
In the limit $\eta \rightarrow 0$, the terms involving the diagonal velocity matrices and the energy denominators in $j_{q,pp}^{CP-\rm{diag}}$ can be recast into:
\begin{align}
    \label{eqn:v_rewrite}
    &-
    \frac{ v^q_{nn} - v^q_{ll}}
    { (\varepsilon_{n\bmk} - \varepsilon_{l\bmk} + \hbar \Omega)^2 }
    =
    \frac{\partial}{\partial k_q} 
    \left(
    \frac{1}{ \varepsilon_{n\bmk} - \varepsilon_{l\bmk} + \hbar \Omega }
    \right).
\end{align}
Thus, combining $j_{q,pp}^{CP-\rm{diag}}$ and $j_{pp}^{q,CP-sh}$, we can get a new term, which we call Fermi~surface~response:
\begin{align}
    \label{eqn:fermi_final}
    &j^{q,\rm{Fermi}}
    =
    -E_r E_s \frac{e^3}{\omega^2} \sum_{n,l} \int d\bmk
    \frac{\partial}{\partial k_q}  (f_{n\bmk} - f_{l\bmk})
    \nonumber \\
    &\times
    \pv{ \left(
    \frac{1}{\varepsilon_{n\bmk} - \varepsilon_{l\bmk} - \hbar \omega}
    -\frac{1}{\varepsilon_{n\bmk} - \varepsilon_{l\bmk} + \hbar \omega}
    \right)
    }
    \Im[v^s_{nl} v^r_{ln}],    
\end{align}
where integration by part has been used and the boundary terms will be zero since the Brillouin zone is a closed manifold.

For a semiconductor, the Fermi-Dirac distribution function ($f_{n\bmk}$ and $f_{l\bmk}$) will be either 0 or 1 and constant throughout the Brillouin zone, so their derivatives over $\bmk$ will be zero.
As a result, $j^{q,\rm{Fermi}}$ will vanish in semiconductors, meaning that no sub-bandgap shift current will occur. \cite{Belinicher82p649}
On the other hand, it can be seen that $j^{q,\rm{Fermi}}$ will survive for a metal where the bands crossing the Fermi surface will be partially occupied, so that $f_{n\bmk}$ and $f_{l\bmk}$ can indeed have non-zero derivative with $\bmk$.
This is another contribution to BPVE, in addition to the circular injection current, and $\omega j^{q,\rm{Fermi}}$ can be shown to be quantized as well for a single Weyl node if $\omega$ is much smaller than the separation between the crossing point and the Fermi level.
But in contrast to the circular injection current which has an energy selection rule, $j^{q,\rm{Fermi}}$ originates from the electronic excitation at Fermi surface and will always contribute to the current regardless of the frequency of light.
Thus, it can be regarded as a non-resonant contribution.
Moreover, the Fermi surface contribution is an intrinsic mechanism similar to shift current in the sense that it is independent of the scattering time and thus insensitive to impurities. 

Until now, we have covered all the major contributions to BPVE for linearly- and circularly-polarized light, from the perspective of time-dependent perturbation theory.
All the mechanisms are second-order in the electric field of light, so they can explain the linear scaling of the experimental photocurrent with light intensity. 
The more exotic \textit{photon-drag effect} by considering the non-vertical optical transitions (non-zero momentum carried by light) has also been discussed by several authors\cite{Grinberg88p87, Fridkin01p654, Mciver12p96}.
Moreover, there could be higher order contributions, such as the \textit{jerk current} originating from the third-order response to electric field (though it is essentially discussing second-order response to the oscillating electric field from light and first-order response to a co-existing static electric field) \cite{Fregoso18p176604}.
Note that the co-existence of static and oscillating electric field is also implicitly considered in the works \cite{Schankler21p1244, Gong16p2879} where the atomic displacements can be driven by a static electric field, which would further influence the BPVE.
Readers who are interested in these mechanisms are encouraged to refer to the original literature cited therein.

\subsection{\label{sec:theory_floquet}Floquet Theory of BPVE}
The theories presented in the last section are all based on time-dependent perturbation theory by treating the optical field as a small perturbation. 
However, recently a different BPVE formalism has been formulated from the Floquet theory, where the optical field is not necessarily weak.
The benefit of the Floquet theory is not immediately obvious, but when combined with non-equilibrium transport theory, it can be easily adapted to investigate the BPVE in a finite system with explicit attachment of electrodes as well as randomly-distributed disorder \cite{Morimoto16pe1501524, Ishizuka17p033015, Ishizuka21p10}.
Here, we outline the framework of the Floquet theory of BPVE in this section and demonstrate that it can lead to the same shift current expression for linearly-polarized light. 
In the next section, we will review its application in finite systems such as Anderson insulators.

We start by giving a brief review of Floquet theory. 
For a general quantum system whose state is described by a state vector $\ket{\Psi(t)}$, its dynamics is determined by the time-dependent Schrodinger equation \cite{Tsuji08p235124,Aoki14p779}:
\begin{align}
    \label{eqn:floquet_tdse}
    \hat{H}(t) \ket{\Psi(t)} = i\frac{d}{dt} \ket{\Psi(t)}.
\end{align}
The Floquet theorem states that for a periodic Hamiltonian:
\begin{align}
    \label{eqn:floquet_periodic_H}
    \hat{H}(t+T) = \hat{H}(t),
\end{align}
where $T$ is the periodicity, there exists a solution in the following form:
\begin{align}
    \label{eqn:floquet_formal_solution}
    \ket{\Psi(t)} = 
    e^{-i\varepsilon_{\alpha}t} \ket{u_{\alpha}(t)}.
\end{align}
Here, $\varepsilon_{\alpha}$ is some quasi-energy that must to be solved for,  $\ket{u_{\alpha}(t)}$ is a function periodic in time: $\ket{u_{\alpha}(t+T)} = \ket{u_{\alpha}(t)}$, and $\{ \alpha \}$ is a set of parameters that label different solutions. 
This is reminiscent of the Bloch theorem which states that for a potential periodic in real space, $\hat{V}(\bmr + \bmrr) = \hat{V}(\bmr)$, the solutions of the time-independent Schrodinger equation must have the form $\ket{\psi(\bmr)} = e^{i\bmk \cdot \bmr} \ket{u(\bmr)}$, where $\ket{u(\bmr)}$ is a lattice-periodic function.

Since $\ket{u_{\alpha}(t)}$ has the periodicity $T$, it can be Fourier transformed into: $\ket{u_{\alpha}(t)} = \sum_n e^{-i n \omega t} \ket{u_{\alpha}^n}$, with $\omega=2\pi/T$.
Now, with the definition of Floquet mode $\ket{u_{\alpha}^n}$, Eq.~(\ref{eqn:floquet_tdse}) can be Fourier transformed:
\begin{align}
    \label{eqn:floquet_ft_tdse}
    \sum_n \hat{H}_{mn} \ket{u_{\alpha}^n} 
    =(\varepsilon_{\alpha} + m \omega) \ket{u_{\alpha}^m}, 
\end{align}

\begin{align}
    \label{eqn:floquet_ft_H}
    \hat{H}_{mn} \equiv \frac{1}{T} \int_{-T/2}^{T/2}
    dt e^{i(m-n) \omega t} \hat{H}(t).
\end{align}
One can see that the dimension of the original Hamiltonian $\hat{H}(t)$ has been augmented by the Floquet indices $n$ and $m$, and in fact they span all integers from $-\infty$ to $+\infty$.
However, since each element in the matrix $\hat{H}_{mn}$ corresponds to the transition probability from $n$-th Floquet mode to $m$-th Floquet mode, in practice if one only considers the low-energy excitations from the ground state, then the Floquet indices will be truncated to a small value to reflect the consideration that the higher-order excitations are neglected.
As another note, the Floquet modes are usually unknown, but the matrix $\hat{H}_{mn}$ has explicit forms provided that $\hat{H}(t)$ can be expressed in a known basis, so one must diagonalize $\hat{H}_{mn}$ to find the Floquet modes $\ket{u_{\alpha}^n}$ and the quasi-energies $\varepsilon_{\alpha}$.

To apply the general Floquet theory to a 1D two-band model in the context of optical excitations, the time-dependent Hamiltonian will take the minimal-coupling form (within the dipole approximation):
\begin{align}
    \label{eqn:floquet_twoband_H}
    \hat{H}(t) = \hat{H}_0 + e \vop \cdot \hat{A}(t),
\end{align}
$\hat{H}_0$ being the Hamiltonian for the two-band model that is time-independent and already diagonalized. 
By Fourier transforming Eq.~(\ref{eqn:floquet_twoband_H}) and focusing on two specific Floquet modes, one being the conduction band with Floquet index $n=0$ and the other being the valence band with Floquet index $n=-1$ (meaning that the valence band is excited and dressed by one photon), we can get a $\mathrm{2\times2}$ Floquet Hamiltonian $H^F_{mn} \equiv H_{mn} - m\omega$ :
\begin{align}
    \label{eqn:floquet_twoband_floquet}
    H^{F}=\left(
    \begin{array}{cc}
    \varepsilon_{v}^{0}+\omega 
    & -i e v_{vc}{A}^{0}  \\
    i e v_{cv}{A}^{0}            
    & \varepsilon_{c}^{0}
    \end{array}
    \right),
\end{align}
where $\varepsilon_{v}^{0}$ and $\varepsilon_{c}^{0}$ are the band energies for the original two-band model, and $v_{vc}$ and $v_{cv}$ are the off-diagonal velocity matrix elements\cite{Morimoto16pe1501524}. 
One can visualize the diagonal elements of $H^F$ as a valence band shifted up by $\omega$ and an original conduction band, and where they are crossing each other, transitions may happen.
The off-diagonal elements of $H^F$ will couple the photon-dressed bands explicitly and lift the degeneracy at the crossing points to form Floquet modes. 

One way to compute the current using the Floquet Hamiltonian is to use the definition of velocity operator for a driven system:
\begin{align}
    \label{eqn:floquet_v_op}
    \tilde{{v}} = \frac{1}{\hbar} 
    \frac{\partial H^F} {\partial k},
\end{align}
and then compute the current from the density operator as in Eq.~(\ref{eqn:current_density_op}), but now the density operator is also dressed by photons in a similar way as Eq.~(\ref{eqn:floquet_ft_H}) \cite{Morimoto16pe1501524}.
We reserve the detailed discussion of how to obtain the dressed density operator to the next section, but the result after some algebra will take the following form:
\begin{align}
    \label{eqn:floquet_final_current}
    j  
    &= \frac{\pi e^3}{\omega^2} E^2  \int d k
    |\braket{vk | \hat{v} | ck}|^2
    \nonumber \\
    & \times
    \left(-\frac{\partial \phi_{v c}(k,k)}{\partial k}-\left[\chi_{c} 
    (k)-\chi_{v}(k)\right]\right) 
    \delta\left(\varepsilon_{vk}-{\varepsilon_{ck}}
    +\hbar \omega\right).
\end{align}
Comparing Eq.~(\ref{eqn:floquet_final_current}) with Eq.~(\ref{eqn:shift_current}), it is obvious that they are equivalent for a 1D two-band model, and Eq.~(\ref{eqn:floquet_final_current}) from Floquet theory can be generalized into Eq.~(\ref{eqn:shift_current}) by performing the same treatment for all the pairs of bands involved in the optical excitation.

\begin{figure}
    \centering
    \includegraphics[width=0.45\textwidth]{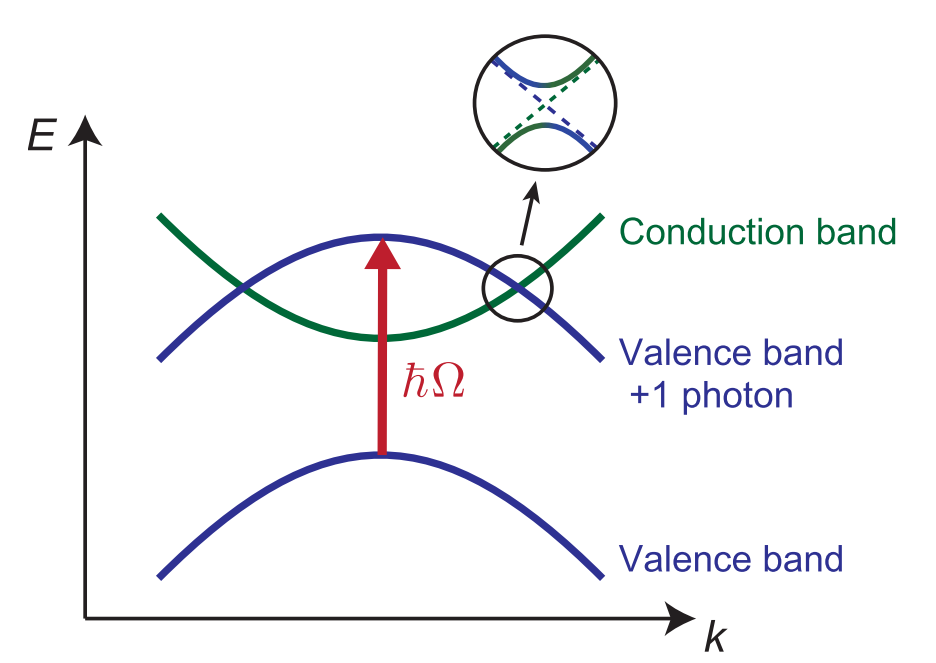}
    \caption{Floquet bands.
    Under the drive of monochromatic light, energy bands evolve into Floquet bands,which describe Bloch states dressed with photons. When two Floquet bands cross, the interaction due to the electron-light coupling will open up a gap and will show an anticrossing. 
    The lower-order nonlinear optics will happen between the anti-crossed Floquet bands.
    Reproduced from the work \cite{Morimoto16pe1501524}.
    }
    \label{fig:floquet_bands}
\end{figure}

\subsection{\label{sec:theory_finte_sys}Finite Systems}
Seeing the similarity between Floquet theory and perturbation theory for BPVE, readers may wonder what advantage the Floquet theory could offer over the perturbation theory.
In this subsection, we show an important application of Floquet BPVE theory, which is the computation of photocurrent for a finite system. 
Going beyond the Floquet theory, it is also possible to investigate the temporal behavior of BPVE away from the steady-state by a more general non-equilibrium transport theory.

To simplify the discussion, most works are based on (but not limited to) the 1D Rice-Mele model \cite{Rice82p1455}, whose Hamiltonian in real space can be written as:
\begin{align}
    \label{eqn:finite_RM_model}
    \hat{H}^{RM} = -\sum_{j} \left(\delta + (-1)^j \frac{B}{2} \right)
    \left(\crea_{j+1} \anni_{j} + \crea_{j} \anni_{j+1} \right)
    +(-1)^j\frac{d}{2},
\end{align}
where $\delta + (-1)^j \frac{B}{2}$ is the staggered hopping parameter, $(-1)^j\frac{d}{2}$ is the staggered on-site energy, and $\crea_j$ and $\anni_j$ are the creation and annihilation operators for electrons in the Rice-Mele system.
This geometry will break the inversion symmetry, and removing either the staggered hopping or the staggered onsite energy will recover the inversion center.
Then, to take into account the influence of photo-excitation, the vector field of light can be incorporated via the Peierls substitution:
\begin{align}
    \label{eqn:finite_H_A}
    \hat{H}^{A} = -i \calla \sin(\omega t) 
    \sum_j \left(
    \delta + (-1)^j \frac{B}{2}
    \right)
    \left(\crea_{j+1} \anni_{j} - \crea_{j} \anni_{j+1} \right),
\end{align}
where $\calla$ is a parameter that characterizes the strength of the vector field, so it is proportional to $A_0$ in Eq.~(\ref{eqn:e-light}) \cite{Ishizuka17p033015}.
In addition, as we are considering explicit attachment of electrodes to a finite system, two leads are placed at the left and right end, whose Hamiltonians and their couplings to the Rice-Mele model are denoted by: 
\begin{align}
    \label{eqn:finite_H_leads}
    H^{X} = \sum_{n} \cread_{nX} \annid_{nX},~X = L,~R,
\end{align}
\begin{align}
    \label{eqn:finite_RM_leads_couple}
    H^{C} = \sum_{n, X, j} 
    \left(V_{nX, j}\cread_{nX} \anni_j
    + V^*_{nX, j}\crea_j \annid_{nX}
    \right).
\end{align}
Here, $\cread_{nX}$ and $\annid_{nX}$ are the creation and annihilation operators for the electrons in the leads, and $V_{nX, j}$ is the coupling strength between the electronic states in Rice-Mele system and the electronic states in the leads \cite{Meir96p5265, Jishi13Feynman}.
Now that the model has been set up, it can be used to investigate more specific situations.

\subsubsection{Local Photoexcitation}
\label{sec:local_photo}
For the local photoexcitation, the summation over $j$ in Eq.~(\ref{eqn:finite_H_A}) will be restricted to a certain range so that only portion of the Rice-Mele system is irradiated.
Then, to compute the photocurrent from local excitation, we can compute the particle change rate in the leads as the generated current will be transported through them.
Following this idea using the concept of non-equilibrium Green's function, one can first arrive at the famous Meir-Wingreen formula for a non-driven system, which will be extended to a light-driven system later \cite{Meir96p5265, Jishi13Feynman}:
\begin{align}
    \label{eqn:finite_meir_wingreen}
    j = &\frac{ie}{4\pi \hbar} \int_{-\infty}^{\infty} d\tilde{\omega} 
    \Tr
    \Big\{
    \left[
    f_L(\tilde{\omega})\Gamma^L(\tilde{\omega}) - f_R(\tilde{\omega})\Gamma^R(\tilde{\omega})
    \right]
    \nonumber \\
    &\times
    \left[
    G^R(\tilde{\omega}) - G^A(\tilde{\omega})
    \right]
    +\left[
    \Gamma^L(\tilde{\omega}) - \Gamma^R(\tilde{\omega})
    \right]
    G^{<}(\tilde{\omega})
    \Big\}.
\end{align}
Here, $f_L$ and $f_R$ are the Fermi-Dirac distribution functions for electrons in left and right leads, respectively, and $G^R$, $G^A$, $G^<$ are the retarded, advanced, and lesser Green's function for the Rice-Mele system \cite{Jishi13Feynman,Rammer86p323}.
$\Gamma^L$ and $\Gamma^R$ are called \textit{level-width functions} characterizing the coupling between the leads and the Rice-Mele system; some reasonable approximation, such as the wide-band approximation where $\Gamma^L=\Gamma^R=\Gamma$, can be used to treat them \cite{Ishizuka17p033015, Meir96p5265}.
Note that we use $\tilde{\omega}$ as the intermediate variable, which will be integrated over, so it is different from the light frequency $\omega$.

To obtain numerical values for $G^R$, $G^A$, and $G^<$, one can group them in a $2\times2$ matrix (Keldysh space) \cite{Rammer86p323, Tsuji08p235124},
\begin{align}
    \label{eqn:finite_keldysh_function}
    G = 
    \left(
    \begin{matrix}
    G^R & G^K\\
    0 & G^A
    \end{matrix}
    \right),
\end{align}
with $G^K \equiv G^R-G^A+2G^<$,
and solve the equation of motion in the frequency space for $G$,
\begin{align}
    \label{eqn:finite_eom_non_driven}
    &\left(\tilde{\omega}  - H^{RM}- \Sigma(\tilde{\omega})\right) G(\tilde{\omega}) = 1,
    \\
    \label{eqn:finite_self_energy}
    &\Sigma_{jk}(\tilde{\omega}) = \sum_{n\alpha} V_{j,n\alpha} [G_0]_{n\alpha,n\alpha}(\tilde{\omega}) V_{n\alpha,k},
\end{align}
where $G_0$ is the Green's function for the non-interacting leads \cite{Ishizuka17p033015}.
It is within the self-energy $\Sigma(\tilde{\omega})$ where approximations can be made for $\Gamma^R$ and $\Gamma^L$.
We treat the coupling to the leads as a perturbation to the central Rice-Mele system, and the information about this coupling is encapsulated in the self-energy $\Sigma(\tilde{\omega})$.
Note that the equation of motion must be solved for each frequency $\tilde{\omega}$ from $-\infty$ to $+\infty$, but in practice a discrete $\tilde{\omega}$-grid is used, and the frequency range is truncated to a point where the integral in Eq.~(\ref{eqn:finite_meir_wingreen}) is sufficiently converged.

As promised, we can extend the above scheme into light-driven systems, from which we can compute the photocurrent \cite{Tsuji08p235124, Aoki14p779}.
The extension needs to use the Green's function in Floquet representation.
Similar to the definition of Floquet Hamiltonian Eq.~(\ref{eqn:floquet_ft_H}), we can first Fourier transform $G$ in time-space to the \textit{Wigner representation}:
\begin{align}
    \label{eqn:finite_wigner}
    G_{\kappa}(\tilde{\omega}) = \frac{1}{T}
    \int_{-\infty}^{+\infty} dt_r 
    \int_{-T/2}^{T/2} dt_a  
    e^{i\tilde{\omega} t_r + i\kappa \omega t_a} G(t, t'),
\end{align}
where ${t_r \equiv t - t'}$ and ${t_a \equiv (t + t') / 2}$. 
Then, the Floquet representation $G_{\kappa \kappa'}(\tilde{\omega})$ is defined as
\begin{align}
    \label{eqn:finite_floquet_rep}
    G_{\kappa \kappa'}(\tilde{\omega}) \equiv G_{\kappa - \kappa'} 
    \left(
    \tilde{\omega} + \frac{\kappa + \kappa'}{2}\omega
    \right).
\end{align}
With the definition of Floquet representation, Eqs.~\ref{eqn:finite_meir_wingreen}-\ref{eqn:finite_self_energy} can be easily modified to the Floquet
space, and the modified Eq.~(\ref{eqn:finite_meir_wingreen}) can give us the photocurrent for local photoexcitation. 
For more details of how the modification is done, see the works \cite{Ishizuka17p033015, Tsuji08p235124}.

\begin{figure*}
    \centering
    \includegraphics[width=1\textwidth]{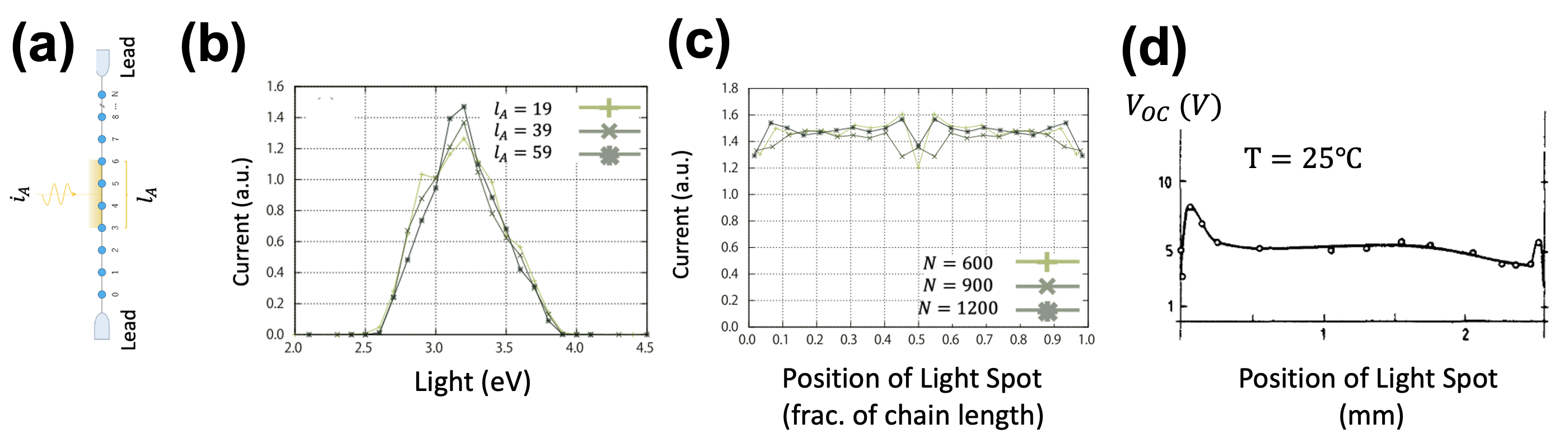}
    \caption{Local photoexcitation. 
    (a) The model system under study. It is composed of a long Rice-Mele chain attached by two leads, and only a finite section of it is illuminated by light. 
    (b) The computed photocurrent from the Floquet-Green's function approach and its wavelength dependence. Different traces correspond to different widths of the light spot. 
    The photocurrent is insensitive to the width of light illumination.
    (c) The dependence of the photocurrent on the position of the light spot.
    Different traces correspond to different lengths of the Rice-Mele chain. 
    The photocurrent is insensitive to the light spot.
    (d) Experiments of BPVE open-circuit voltage in \ch{BaTiO3} showing the insensitivity to light spot.
    Ref.~\cite{Ishizuka17p033015} is reproduced with permission from New J. Phys. 19 033015 (2017). Copyright 2017 IOP Publishing Ltd and Deutsche Physikalische Gesellschaft. 
    Ref.~\cite{Koch75p847} is reproduced with permission from Solid State Commun. 17, 7 (1995). Copyright 1975 Published by Elsevier Ltd.}
    \label{fig:local_excitation}
\end{figure*}

The calculated photocurrent from this scheme is shown in Fig.~\ref{fig:local_excitation}. 
The first thing to notice is that its spectral feature can be understood from an analytical expression of shift current for 1D Rice-Mele model with periodic boundary condition, which only has one diverging peak at the band edge \cite{Fregoso18p176604}. 
Since now the extra coupling is included (to leads), the divergent density of states at the band edge is broadened so that we now can observe a major peak with finite magnitude. 
Another feature of the local photoexcitation is the insensitivity of the photocurrent to the location of the irradiation as can be seen in Fig.~\ref{fig:local_excitation}c, and this is argued to be a peculiar feature of shift current since local excitation will excite delocalized valence electrons to delocalized conduction band, where the coordinate shift happens during the transition. 
Thus, due to the delocalization of the wave functions in a periodic system, the coordinate shift is expected to happen coherently through the 1D chain regardless of location in which the photoexcitation happens.
This is in agreement with previous experiments showing the insensitivity of the photocurrent to the irradiation location (Fig.~\ref{fig:local_excitation}d) \cite{Koch75p847}, a feature that can only be captured by the Floquet BPVE theory.

This model can also be used to investigate the influence of disorder in the generation of photocurrent \cite{Ishizuka21p10}.
To this end, a random potential $V_{rnd}$ can be added to the onsite energy $d$ in Eq.~(\ref{eqn:finite_RM_model}) to form the model for Anderson insulator, and the electron-phonon coupling is also taken into account which manifests as another term in the self-energy.
Tuning the magnitude of the random potential, one can change the strength of disorder and observe how photocurrent behaves. 
The simulation results for the Anderson insulator model show that the disorder could greatly localize the wave function such that now the current generation has a strong dependence on the location of local photoexcitation. 
For the excitation happening right at the middle of the 1D chain, the current can be hardly transported to the leads due to the localization effect, whereas getting closer to the ends will generate a larger current.
On the other hand, under the uniform illumination, the existence of photocurrent is rather robust against the disorder. 
Even in the regime of strong disorder, i.e. when the disorder potential is larger than the band width \cite{Ishizuka21p10}, the shift current will still persist with a reduced magnitude. 
The comparison between the local photoexcitation and uniform photoexcitation demonstrates that the BPVE as a bulk property will be robust to the scattering, though the further propagation of the current away from the bulk region will be restricted by the disorder.

\begin{figure}
    \centering
    \includegraphics[width=0.45\textwidth]{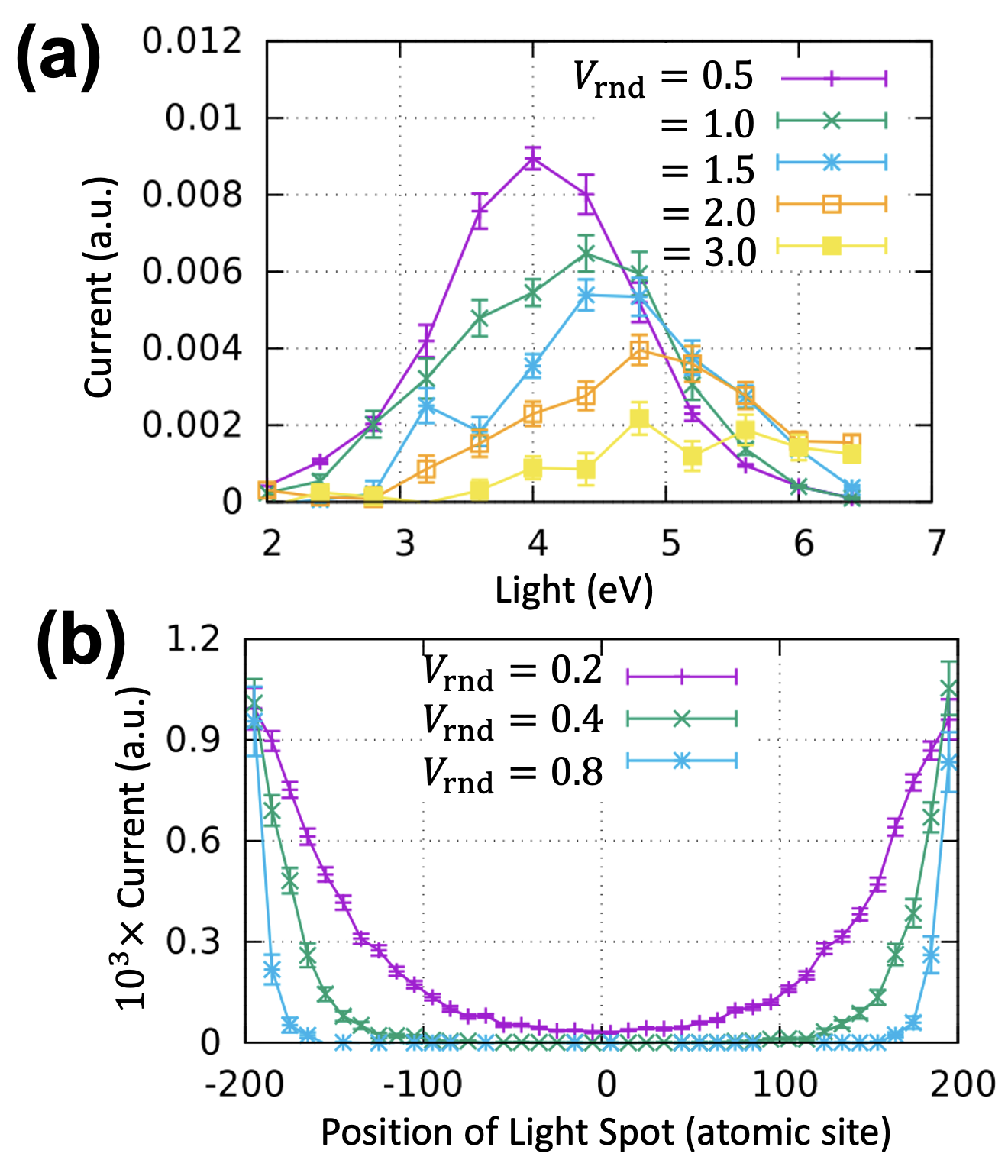}
    \caption{Photocurrent in the Anderson insulator.
    (a) Simulation from the Floquet-NEGF formalism shows that in the presence of disorder, the photocurrent under uniform illumination will be significantly changed, but finite magnitude will persist even for strong disorder. A general trend is that a weaker photocurrent profile corresponds to a stronger disorder. 
    (b) When there is strong disorder, the local photoexcitation will generate a photocurrent that depends on the position of the light spot. When illuminating the central region, the photocurrent cannot propagate to the leads, and a stronger disorder will confine the photocurrent in a broader range. 
    Reproduced from the work \cite{Ishizuka21p10}.
    \label{fig:anderson_insulator}}
\end{figure}

Although the models described here are especially suitable to investigate finite systems, the Floquet-NEGF formalism can also work for extended periodic system as discussed in the last section \cite{Ishizuka17p033015, Morimoto16p245121}. 
To do so, we apply the periodic boundary condition to Eq.~(\ref{eqn:finite_RM_model}) and omit the leads.
Then, instead of considering the coupling to the leads as the self-energy, one can consider that each eigenstate of the periodic $\hat{H}^{RM}$ is coupled to a heat bath (similar to the treatment of electron-phonon coupling in the Anderson insulator model \cite{Ishizuka21p10}).
If we still use the wide-band approximation, i.e., assuming a constant $\Gamma$ for the level-width function, then we can calculate the lesser Green's function $G^<$, from which the density operator can be obtained according to \cite{Jishi13Feynman, Rammer86p323, Bajpai19p025004}
\begin{align}
    \label{eqn:finite_rho_lesser}
    \oprho(t) = -i G^<(t, t).
\end{align}
Combining with Eq.~(\ref{eqn:floquet_v_op}), we can finally arrive at the photocurrent Eq.~(\ref{eqn:floquet_final_current}) in the Floquet BPVE theory.

\subsubsection{Temporal Response}

\begin{figure*}
    \centering
    \includegraphics[width=0.9\textwidth]{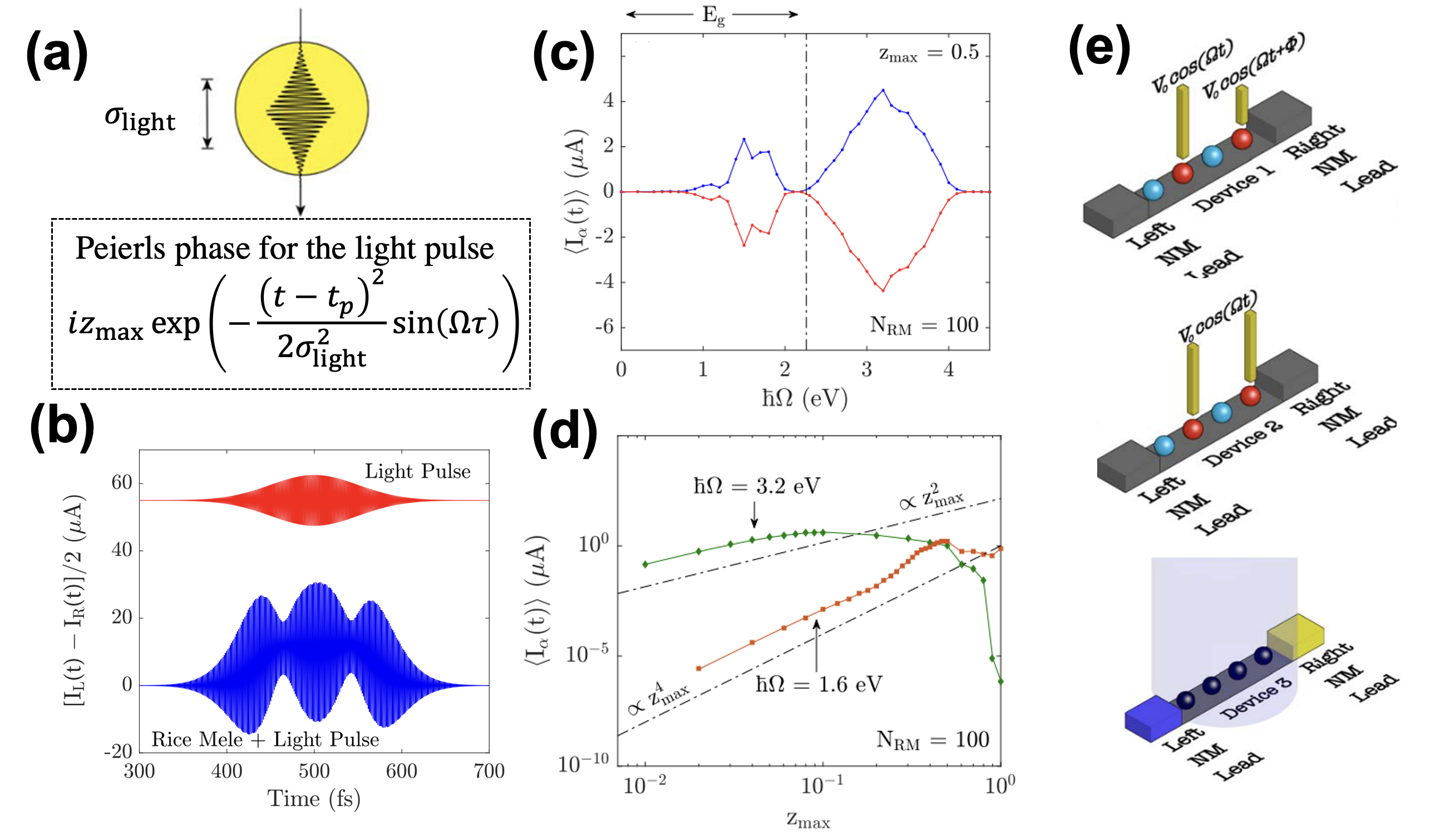}
    \caption{Temporal response. 
    (a) The light pulse used in the numerical simulation.
    (b) The temporal photocurrent (blue) with time, showing linear component as well as more complicated nonlinear component.
    (c) The photocurrent going through each lead (blue: left lead; red: right lead) and their wavelength dependence. It is interesting to observe a sub-bandgap response, which is attributed to a two-photon process. 
    (d) The scaling of the sub-bandgap response and above-bandgap response with the light intensity. 
    The fourth-order scaling of the sub-bandgap response confirms the two-photon process.
    (e) Different numerical experiments that can be performed in the Rice-Mele model. The system can be driven by two light pulses with different phases (left), or the light pulses can be in phase (middle). In addition, the photoresponse of a centrosymmetric system with asymmetric leads can also be modeled (right).
    Reproduced with permission from J. Phys.: Mater. 2 025004 (2019).\cite{Bajpai19p025004} Copyright 2019 The Author(s). Published by IOP Publishing Ltd.}
    \label{fig:temporal_response}
\end{figure*}

Besides the photocurrent at steady-state, one can also study the temporal photocurrent in the finite system Eq.~(\ref{eqn:finite_RM_model}) by propagating the Green's function $G$ in real-time using a quantum-Liouville-like equation \cite{Bajpai19p025004}.
The photocurrent can still be calculated from the particle change rate in the leads, but instead of focusing on the DC component after Fourier transformation, we now can calculate how the current evolves with time.
Besides, this approach is also able to model a laser pulse (Fig.~\ref{fig:temporal_response})a), in contrast to the perturbative BPVE theory which generally assumes a continuous wave light.

When the light pulse has a Gaussian envelope, the photoconductivity can be directly calculated following the procedure in the work \cite{Bajpai19p025004}, and the results are displayed in Fig.~\ref{fig:temporal_response}b, where the linear conductivity is dominant (same periodicity as the light frequency), and the DC component (BPVE) and some higher-order responses (second harmonic generation, for example) can also be observed.
What is more interesting is that if one plots the time-averaged current through each lead as a function of photon frequency, then this time-dependent NEGF approach can obtain a sub-bandgap current, which is clearly at odds with the conclusion from the second-order perturbation theory (Section.~\ref{sec:theory_circular_light}).
However, if one plots the scaling of the sub-bandgap current against the strength of the light field, then the observed quartic scaling indicates that this sub-bandgap current is indeed a fourth-order (two-photon) response. So this sub-bandgap photocurrent is only significant when the light field is stronger than the perturbative regime.

Readers may have noticed that in this section, we have kept using \textit{photocurrent} instead of the more specific mechanism such as the shift current or ballistic to refer to the simulated DC current.
This is due to the fact that what we calculated for the finite system is always the total current, and there is no clear way of separating the current into different mechanisms, as opposed to the perturbative BPVE theory, where the shift current is defined as the off-diagonal contribution of the density operator, whereas the ballistic current and injection current refer to as the diagonal contributions. 
One can argue that for a static 1D Rice-Mele system under linearly-polarized light, the only possible mechanism with second-order scaling is the shift current, but when there is disorder, for instance in the Anderson model \cite{Ishizuka21p10}, scatterings would come into play to give ballistic current, and the approaches developed here do not clearly differentiate the characteristics of the current.
Nevertheless, the Floquet- and NEGF-based BPVE theory for finite systems provides a different perspective for looking at the photocurrent for homogeneous materials, and it enables many interesting numerical experiments which are otherwise difficult to model in the perturbation theory, such as exciting materials with two light sources.

\section{Materials Design}
\label{sec:mat_design}
Shortly after the realization of first-principles calculations of BPVE, especially the less computationally demanding calculation of shift current for real materials, various proposals have been made to further enhance the shift current via numerical materials design \cite{Schankler21p1244, Cook17p14176, Wang17p115147, Gong16p2879, Rangel17p067402,Tan16p237402}. 
Thus, several design rules have been proposed that try to relate the shift current with certain properties of materials, such as the electric polarization, the bonding character, and the delocalization of wave functions \cite{Tan16p16026, Tan19p084002, Young13thesis}.
On the other hand, it is of interest to know what the upper limit of the shift current is so that the maximum efficiency can be predicted \cite{Tan19p085102}.
In addition, there exists another material design strategy which is to exhaust all possibilities in the parameter space to maximize the shift current \cite{Cook17p14176, Schankler21p1244}. 
It is our goal in this section to survey these endeavors in materials design to enhance the shift current.

\subsection{Influencing factors}
\begin{figure}
    \centering
    \includegraphics[width=0.8\textwidth, angle=270]{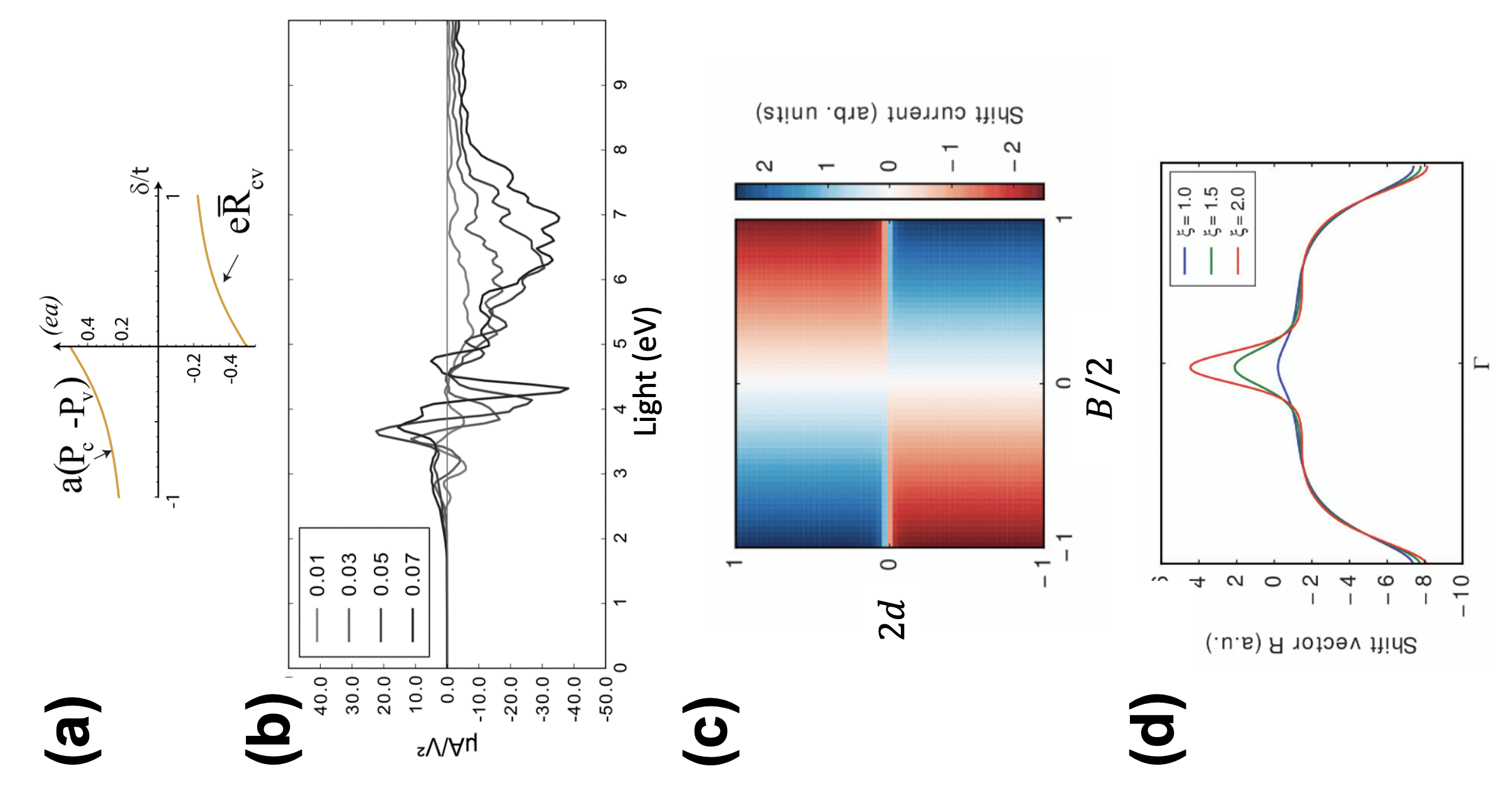}
    \caption{Influencing factors for shift current.
    (a) The correlation between the polarization difference and the integrated shift vector for a two-band model with a simplified shift current expression. Under these conditions, the integrated shift vector is proportional to the polarization difference.
    (b) A more realistic material (\ch{PbTiO3}) shows that in reality, the relation between the ground state electronic polarization and the shift current is complicated, without clear rules.
    (c) The shift current response of the Rice-Mele model. The shift current is shown as a function of the hopping asymmetry and the on-site energy asymmetry. Negative values of shift current (red) correspond to a current flowing to the left, and positive values of shift current (blue) correspond to a current flowing to the right.
    (d) The dependence of shift vector on the delocalization of the wave function. A larger delocalization corresponds to a larger shift vector.
    Ref.~\cite{Fregoso17p075421} is reproduced with permission from Phys. Rev. B 96, 075421 (2017). Copyright 2017 American Physical Society. 
    Ref.~\cite{Tan16p16026} is reproduced with permission from Npj Comput. Mater. 2, 16026 (2016). Copyright 2016 Springer Nature Limited}
    \label{fig:influence_factor}
\end{figure}

Given the appearance of the Berry connection in the expression Eq.~(\ref{eqn:shift_current}), it is tempting to relate the magnitude of shift current with the electronic polarization, which takes the following form (for the simplest non-degenerate 1D situation) \cite{Vanderbilt00Berry}: 
\begin{align}
    \label{eqn:design_epol}
    P = \sum_n \int{dk \chi_n(k)},
\end{align}
where the index $n$ runs over all the occupied bands. 
However, as the shift current involves the excitation to conduction bands, the relation between shift current and electronic polarization is elusive. 
In particular, for a system that is non-centrosymmetric but non-polar, such as \ch{GaAs}, the shift current can still appear. 
Therefore, by this simple analysis electric polarization may not be a good measure to determine the magnitude of BPVE.
Nonetheless, Fregoso \cite{Fregoso18p176604} pointed out that at least in 2D, the integrated shift current over all frequency can be understood from the polarization \textit{difference} for valence band and conduction band. 
What is found is that for a two-band model, when making the drastic approximation that the transition intensity is constant, the integrated shift current is proportional to the shift vector integrated over the Brillouin zone, which is equal to the polarization difference between the valence band and conduction band (Fig.~\ref{fig:influence_factor}a). 
Thus, qualitatively speaking, the magnitude of shift current can be inferred from the polarization difference, and this further shows that ground state electronic polarization alone cannot be used to predict the shift current. 

First-principles simulations of \ch{PbTiO3} demonstrate the same point.
As the polarization of \ch{PbTiO3} can be modified by the displacement of oxygen octahedra, it is instructive to make a series of displacements starting from the paraelectric structure and calculate the corresponding shift current \cite{Young12p116601}. 
It can been seen from Fig.~\ref{fig:influence_factor}b that no systematic change has been found for shift current with different oxygen displacement. 
In particular, when the oxygen octahedron displacement is of the 0.01 (of the lattice parameter) amplitude, the shift current at 3.2~eV above the band gap is negative, which will further reverse its sign when the displacement amplitude is increased to 0.07. 
Meanwhile, there is complicated peak shifting as the displacement increases. 
Thus, for real materials, polarization is not a straightforward metric to estimate shift current.

Another possible influencing factor is the bonding character, which can be best illustrated via a two-band model, specifically the Rice-Mele model Eq.~(\ref{eqn:finite_RM_leads_couple}).
It is noted that both the onsite energy and the hopping parameter can control the polarization, so one should not equate the bonding character with the polarization.
By varying the hopping parameter and the onsite energy, the shift current can be computed for each combination of the parameters, as shown in Fig.~\ref{fig:influence_factor}c. 
The first observation is that the shift current is insensitive to the the asymmetry of the onsite energy, even though the polarization can be effectively modified by changing the onsite energy. 
This is again consistent with the previous argument that polarization is not a good metric for shift current. 
On the other hand, changing the bonding character, i.e., changing the hopping parameter, will strongly affect the shift current. 
Specifically, the more asymmetric the bonding character is, the larger shift current we can get from the Rice-Mele model.
Therefore, this model demonstrates the direct correlation between the bonding character and the shift current.

The third factor that has a direct impact on shift current is the extent of the delocalization of the wave function \cite{Tan19p084002, Young13thesis}.
To demonstrate this, we go back to the Rice-Mele model, but this time, the hopping is no longer restricted to the nearest neighbors. 
Hoppings to farther away neighbors are assumed to decay exponentially and can be represented by $e^{-R/\xi}$, where $R$ is the hopping distance and $\xi$ characterizes the decay rate. 
Thus, a slow decay of the hopping parameter will generate a more delocalized wave function.
After calculating the shift vector for different $\xi$, it can be seen that more delocalized wave functions generate a larger shift vector, and thus a larger shift current.

In a short summary of this subsection, three influencing factors for shift current have been identified, which are the electric polarization, bonding character, and the delocalization of wave functions, but their relations to shift current are mostly correlative, and no definite overarching rules have been set.
Nevertheless, these factors can be the first targets to be optimized in an attempt to enhance the shift current.  

\subsection{Upper bound}
Based on identified factors, such as the bonding asymmetry and the wave function delocalization, that can increase the shift current, a natural question occurs: what is the maximum shift current that can be attained by optimizing these factors?
Tan \textit{et al.} \cite{Tan19p085102} investigated this problem by trying to formulate an expression for the integrated shift current over all frequencies, which is more relevant for photovoltaic applications.
With a generic two-band model, the integrated shift current would have a rather simple form:
\begin{align}
    \label{eqn:design_generic_2band}
    H(\bmk) = \mathbf{h}(\bmk) \cdot \boldsymbol{\tau},
\end{align}
\begin{align}
    \label{eqn:design_inte_sc}
    \int{\sigma^{q,sh}_{rs} d\omega} 
    =
    \frac{\pi e^3}{2 \hbar} 
    \abs{\int{d\bmk} 
    \frac{\mathbf{h}(\bmk) \cdot \mathbf{h'}(\bmk)\times \mathbf{h''}(\bmk)} 
    {2 \abs{\mathbf{h}(\bmk)}} }.
\end{align}
Here, $\boldsymbol{\tau}$ is the vector of three Pauli matrices representing the band degree of freedom, and $\mathbf{h}({\bmk})$ are the parameters for the generic two-band Hamiltonian. The prime and double prime represent the first and second derivative with $\bmk$ along light polarization and current direction, respectively. 
From Eq.~(\ref{eqn:design_inte_sc}), one can see that it is $\mathbf{h}({\bmk})$ that determines the upper limit of the integrated shift current. 
If we assume that the Hamiltonian in real-space decays exponentially with distance, which is often the case (also consistent with the previous Section), then it can be shown that the integrated shift current will have a upper bound that is expressed as a few parameters:
\begin{align}
    \label{eqn:design_upper_bound}
    \int{\sigma^{q,sh}_{rs} d\omega} 
    <
    \frac{2 \pi e^3}{\hbar} \left( \frac{A}{E_g}\right)
    \Xi( \{\bm{R}\}, \boldsymbol{\xi}, \bm{u} ),
\end{align}
where $A$ is the overall hopping amplitude, $E_g$ is the band gap, $\{\bm{R}\}$ denotes the lattice vectors for a given system, $\boldsymbol{\xi}$ characterizes the hopping range along lattice vector directions, and $\mathbf{u}$ the direction along which the current is measured.
$\Xi$ can be understood as a geometric factor, and combined with $A$, it can be used to determine the upper limit of the integrated shift current for systems with a given band gap.

\begin{figure}
    \centering
    \includegraphics[width=0.45\textwidth]{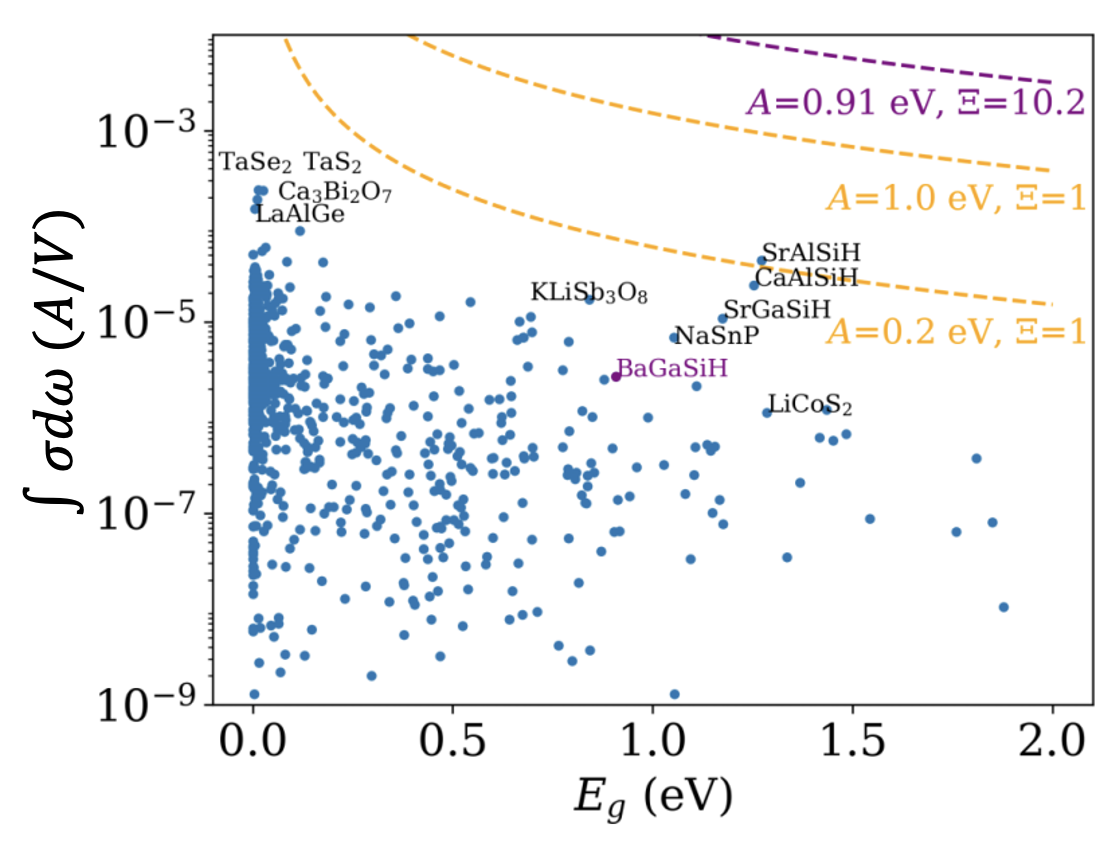}
    \caption{Upper limit of shift current.
    Integrated shift current for a set of semiconductors and semimetals. The largest tensor component of the integrated shift current for each material is plotted against the band gap. Dashed lines indicate the value of the shift current upper bound as a function of the band gap, for different values of hopping strength ($A$) and geometrical factor ($\Xi$)
    Reproduced with permission from Phys. Rev. B 100, 085102  (2019). \cite{Tan19p085102} Copyright 2019 American Physical Society. }
    \label{fig:upper_bound}
\end{figure}

Fig.~\ref{fig:upper_bound} shows the comparison of theoretical upper bound with \textit{ab initio} computed shift current for a database of non-centrosymmetric materials. 
It is remarkable that most materials are well below the theoretical upper limit given by $A=0.2~\rm{eV}$ and $\Xi=1$, and their integrated shift currents closely track the upper limit contour as a function of band gap $E_g$.
For some Zintl-type materials such as \ch{SrAlSiH} (which is above the $A=0.2~\rm{eV},~\Xi=1$ contour), their corresponding $A$ and $\Xi$ values are computed by fitting the DFT band structures to tight-binding models. 
If the band structure of \ch{BaGaSiH} is used, then the corresponding parameters would be $A=0.91~\rm{eV}$ and $\Xi=10.2$, whose upper-bound contour is far above the calculated shift current of Zintl-type materials.

The upper limit of shift current introduced in this section can be used to perform a preliminary screening to select materials that will exhibit large shift current, for which further materials engineering can be done to enhance the photocurrent by considering optimizing the influencing factors as mentioned the last section, or by applying the exhaustive search strategy discussed below. 

\subsection{Exhaustive materials design}
As the influencing factors for shift current are mostly correlative, simply optimizing them would not guarantee the largest shift current.
Therefore, a different strategy has been proposed, which is to go through all the possible values of the parameters in a specified design space and pick the combination that would give rise to the maximum shift current.

\begin{figure*}
    \centering
    \includegraphics[width=1\textwidth]{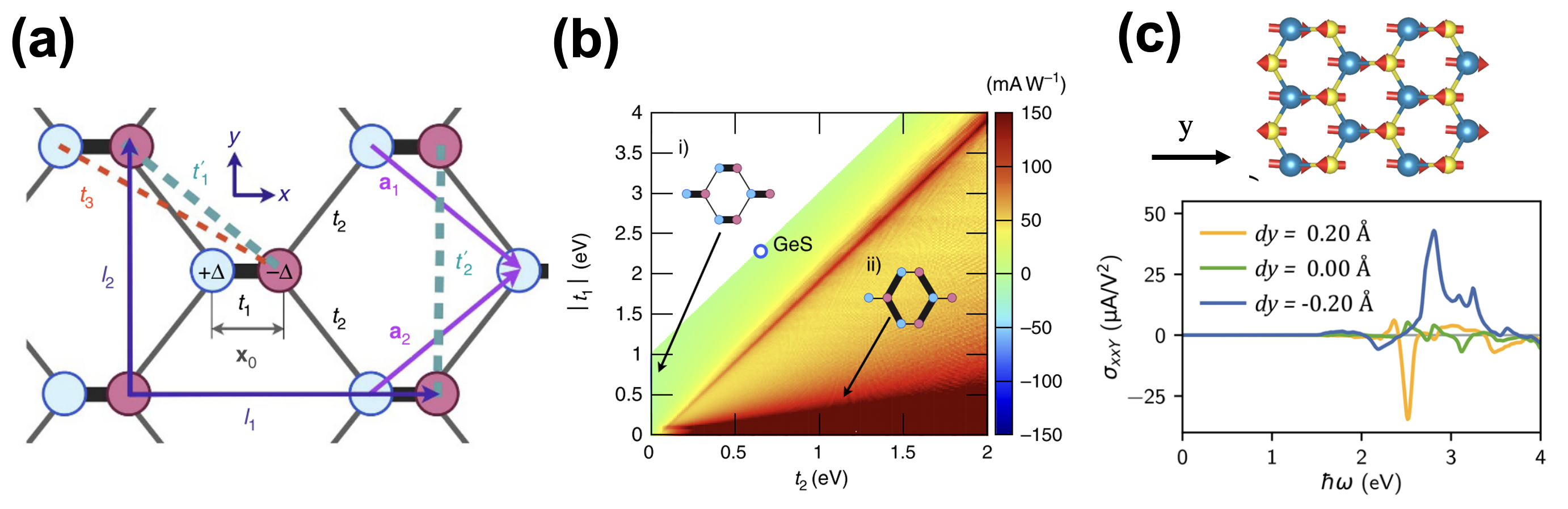}
    \caption{Optimize shift current by exhausting all possibilities in the material design space.
    (a) The tight-binding model used to parametrize the monolayer monochalcogenides. 
    (b) Band-edge shift current for different choices of the parameters in the tight-binding model. Specifically, $|t_1|$, $t_2$, and $\Delta$ are varied (with a fixed band gap as 1.89~eV). $x_0=0.52~\AA$ and $t_3=0$.
    (c) The displacement pattern (upper panel) that can optimize the integrated shift current for \ch{MoS2}, which is obtained by a gradient descent approach, and the enhanced shift current (lower panel) according to the pattern.
    Ref.~\cite{Fregoso17p075421} is reproduced with permission from Nat. Commun. 8, 14176 (2017). Copyright 2017 Springer Nature Limited. 
    Ref.~\cite{Tan16p16026} is reproduced with permission from J. Phys. Chem. Lett. 12, 4 (2021). Copyright 2021 American Chemical Society}
    \label{fig:optimize_sc}
\end{figure*}

The first attempt that adopts this strategy is to express the shift current at the band edge frequency as an analytical expression \cite{Cook17p14176}.
The restriction to band edge response enables the use of a gapped Dirac Hamiltonian, which makes the shift current rather simple to formulate and evaluate. 
For example, the band structures of monolayer monochalcogenides can be parametrized by a two-band model, which can be further simplified if we expand the two-band model around the $\bmk$ point corresponding to the band edge states. 
Then, the band edge shift current only depends on the parameters shown in Fig.~\ref{fig:optimize_sc}a.
By changing the values of these parameters, such as $|t_1|$, $t_2$, and $\Delta$ (while keeping the band gap fixed), the corresponding shift current can be calculated, making it easy to locate the best combination of parameters that optimize the current.
The naturally occuring \ch{GeS}, an example of monochalcogenide, is located in the white circle shown in Fig.~\ref{fig:optimize_sc}b, which shows a huge room for further improvement, as the maximum shift current can be five times larger.

While this approach of simplifying the shift current expression to a few parameters looks promising since it can pinpoint the critical points in the parameter space, it is not very clear how to achieve such parameters for real materials. 
Furthermore, such simplifications have only been done for the band edge, and extending it to optimize the shift current for a broader frequency range is not straightforward. 
Thus, it is desirable to optimize the shift current with respect to some experimentally controllable parameters.
Atomic displacements can play this role, as they can be controlled via static electric field or strain, so examining how shift current changes with different atomic displacements will provide experimental guidance.
To this end, Schankler \textit{et al.}\cite{Schankler21p1244} conducted a systematic study on monolayer \ch{MoS2} by using a gradient descent approach to find the best displacement pattern and magnitudes.
Noticing the smooth change of shift current with the atomic position, they started with the DFT-relaxed (equilibrium) structure and then computed the \textit{change} of the integrated shift current for different local atomic displacements. 
After finding the local displacement pattern that corresponds to the steepest change of the integrated shift current, the structure is distorted according to this pattern, and the same procedure will be repeated until the maximum current is achieved. 

From this gradient descent search, Schankler revealed an extremely simple pattern that can enhance the integrated shift current the most. It is shown in Fig.~\ref{fig:optimize_sc}c, where the \ch{Mo} and \ch{S} have opposite displacements.
This pattern is easily achieved by applying a static electric field since \ch{Mo} and \ch{S} have positive a negative charges, respectively, making them move along opposite directions in response to the electric field. 
The integrated shift current is increased by almost 10 times compared to the equilibrium structure, and the current direction can be flipped by reversing the displacement, i.e., by reversing the direction of the electric field.
Thus, the exhaustive strategy is proved to be very powerful to enhance the shift current, not only in that it can provide insight on what contributes the most to the shift current, but it can also point out a clear route for experimentalists to realize such enhancement.

\section{Summary and Outlook}
In this article, we reviewed the recent development of theories for the bulk photovoltaic effect, including the shift current, ballistic current, injection current, and the Floquet BPVE theory. 
In contrast to the coherent evolution of wave packets in the shift current mechanism, the ballistic current has a more classical nature and originates from the asymmetric carrier generation.
Such asymmetry can be attributed to the additional scattering processes, such as electron-phonon and electron-hole scattering.
Due to the complexity of the appropriate treatment of the additional scattering processes, first-principles calculations of ballistic current have become possible only recently.
On the other hand, the asymmetric carrier generation can also be realized via circularly-polarized light or through magnetization, and such intrinsic ballistic current is usually referred to as the injection current. 
Besides the perturbative approach of investigating BPVE, a non-perturbative Floquet approach is developed and can achieve the same conclusion as the perturbative theory of BPVE for a two-band model. 
Different from the perturbative approach, the Floquet approach can be further extended to study the BPVE for finite systems or even temporal responses. 
We also reviewed some strategies of materials design to enhance the shift current due to its relative simplicity.
Though no overarching rules have been established, several influencing factors have been identified, and the upper limit for shift current has been derived.
Moreover, there exists an exhaustive strategy that can systematically maximize the shift current and provide a viable path to experimental realization.

Despite the inspiring development in the BPVE theory, there are still some opening questions that largely remain unanswered. 
For example, it is still unclear of how an external magnetic field would influence the shift current.
As is well known, the magnetic field can have highly nontrivial influence on the band structure and wave functions of materials, such as giving rise to the famous Hofstadter butterfly energy spectrum and the quantum Hall effect \cite{Thouless82p405, Hofstadter76p2239}, the simple energy-scale argument made by Ivchenko \cite{Ivchenko84p55} needs to be revisited, and a non-perturbative approach of treating the magnetic field when calculating magnetic shift current is preferred.
Second, as also pointed out by the work \cite{Tan16p16026}, the calculation of the open-circuit voltage from first-principles theories has not been demonstrated. 
Since most widely-adopted BPVE theories are based on periodic boundary conditions, there is no explicit termination of the system that resembles the open-circuit condition.
Moreover, the open-circuit voltage also depends on extrinsic properties of materials when computing the total resistivity, such as the concentration of defects or electrode interface properties.
Thus, it is difficult to formulate a theory that reliably captures all the major mechanisms of resistivity. 
The third open question is the role of spin in BPVE. 
All the theories developed so far only treat spin-up and spin-down channels as two separate systems and then compute the photocurrent for each of them.
 (For PT-symmetric systems, each band is still doubly degenerate, making the separation of spin-up and spin-down possible.)
However, the spin-orbital coupling could fundamentally change the expressions derived for shift current and ballistic current because it can cause spin-mixing, or it can be thought of as a perturbation to the electronic structure.
Indeed, a recent work by Rajpurohit \textit{et al.} demonstrates that the spin scattering can cause an extra contribution to the ballistic current \cite{Rajpurohit21arxiv}.
As a result, for systems with strong SOC or non-colinear spins, it is debatable if the BPVE theories developed so far can be directly used to compute their photocurrents.

In summary, the development of BPVE theory allows for a deeper understanding of this phenomenon, and the predictive power embodied by first-principles methods enables the direct comparison with experiments.
In addition, the materials design to enhance the BPVE is also made possible by the newly-developed theories and their numerical implementations.
We hope that our review can provide a perspective on the current stage of the BPVE theory is and encourage the community to extend the exploration of this fruitful field.

\section*{Acknowledgment}
We thank the invaluable discussion with Dr. Lingyuan Gao and Aaron M. Schankler.
This work was supported by the U.S. Department of Energy, Office of Science, Basic Energy Sciences, under Award DE-FG02-07ER46431.


\section*{References}


\bibliography{rappecites.bib, rappecites_temp.bib}

\end{document}